\documentclass[longauth]{aa} 
\usepackage{xcolor}
\usepackage{support-caption}
\usepackage{subcaption}
\usepackage{threeparttable} % For table notes
\usepackage{rotating}
\usepackage{pdflscape}
\usepackage{booktabs} % For better table rules
\usepackage{graphicx}

\newcommand{\ha}{\ifmmode {\rm H}\alpha \else H$\alpha$\fi}
\newcommand{\hb}{\ifmmode {\rm H}\beta \else H$\beta$\fi}
\newcommand{\lya}{\ifmmode {\rm Ly}\alpha \else Ly$\alpha$\fi}
\newcommand{\pg}{\ifmmode {\rm P}\gamma \else Pa$\gamma$\fi}
\newcommand{\lyb}{\ifmmode {\rm Ly}\beta \else Ly$\beta$\fi}
\newcommand{\lyg}{\ifmmode {\rm Ly}\gamma \else Ly$\gamma$\fi}

\newcommand{\flyc}{\ifmmode \mathrm{f}_\mathrm{esc}\mathrm{(LyC)} \else $\mathrm{f}_\mathrm{esc}\mathrm{(LyC)}$\fi}

\def\ergs{\ifmmode \mathrm{erg\hspace{1mm}s}^{-1} \else erg s$^{-1}$\fi}

\def\micron{\ifmmode \mu\mathrm{m} \else $\mu$m\fi}
\def\msun{\ifmmode \mathrm{M}_{\odot} \else M$_{\odot}$\fi}
\def\msunyr{\ifmmode \mathrm{M}_{\odot} \hspace{1mm}{\rm yr}^{-1} \else $\mathrm{M}_{\odot}$ yr$^{-1}$\fi}
\def\zsun{\ifmmode Z_{\odot} \else Z$_{\odot}$\fi}
\def\lsun{\ifmmode L_{\odot} \else L$_{\odot}$\fi}
\def\mstar{\ifmmode \mathrm{M}_{\star} \else M$_{\star}$\fi}

\usepackage{txfonts}
\newcommand{\orcid}[1]{\href{https://orcid.org/#1}{\includegraphics[width=9pt]{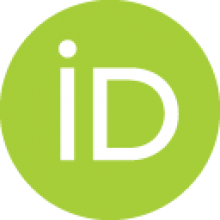}}}
\usepackage{hyperref}
\hypersetup{colorlinks, % = true per default
 linkcolor = blue, 
 urlcolor = blue, 
 citecolor = blue, 
 anchorcolor = blue
}
\usepackage{multicol}
\begin{document} 

\title{The GLASS-JWST Early Release Science Program. IV. Data release of 263 spectra from 245 unique sources}
\subtitle{}

\author{S. Mascia \orcid{0000-0002-9572-7813}\fnmsep\thanks{E-mail: sara.mascia@inaf.it}
 \inst{1, 2}
 \and
 G. Roberts-Borsani \orcid{0000-0002-4140-1367}
 \inst{3}
 \and
 T. Treu \orcid{0000-0002-8460-0390}
 \inst{4}
 \and
 L. Pentericci \orcid{0000-0001-8940-6768}
 \inst{1}
 \and
 W. Chen \orcid{0000-0003-1060-0723}
 \inst{5}
 \and A. Calabrò \orcid{0000-0003-2536-1614}
 \inst{1}
 \and E. Merlin \orcid{0000-0001-6870-8900}
 \inst{1}
 \and D. Paris \orcid{0000-0002-7409-8114}
 \inst{1}
\and P. Santini \orcid{0000-0002-9334-8705}
 \inst{1}
 %%%NIRSPec GLASS Builders
 \and G. Brammer \orcid{0000-0003-2680-005X}
 \inst{6, 7}
 \and A. Henry \orcid{0000-0002-6586-4446}
 \inst{8, 9}
 \and P. L. Kelly \orcid{0000-0003-3142-997X}
 \inst{5}
 \and C. Mason \orcid{0000-0002-3407-1785}
 \inst{6, 7}
 \and T. Morishita \orcid{0000-0002-8512-1404}
 \inst{10}
 \and T. Nanayakkara \orcid{0000-0003-2804-0648}
 \inst{11}
\and N. Roy \orcid{0000-0002-4430-8846}
 \inst{9}
 \and X. Wang \orcid{0000-0002-9373-3865}
 \inst{12, 13, 14}
 \and H. Williams \orcid{0000-0002-1681-0767}
 \inst{5}
 %%%%
 \and K. Boyett \orcid{0000-0003-4109-304X}
 \inst{15, 16}
 \and M. Brada{\v c} \orcid{0000-0001-5984-0395}
 \inst{17, 18}
 \and M. Castellano \orcid{0000-0001-9875-8263}
 \inst{1}
 \and K. Glazebrook \orcid{0000-0002-3254-9044}
 \inst{21}
 \and T. Jones \orcid{0000-0001-5860-3419}
 \inst{18}
\and L. Napolitano \orcid{0000-0002-8951-4408}
 \inst{1}
\and B. Vulcani \orcid{0000-0003-0980-1499}
\inst{19}
\and P. J. Watson \orcid{0000-0003-3108-0624}
\inst{19}
\and L. Yang \orcid{0000-0002-8434-880X}
\inst{20}}
\institute{INAF – Osservatorio Astronomico di Roma, via Frascati 33, 00078, Monteporzio Catone, Italy
 \and 
Dipartimento di Fisica, Università di Roma Tor Vergata, 
Via della Ricerca Scientifica, 1, 00133, Roma, Italy
\and
Department of Astronomy, University of Geneva, Chemin Pegasi
51, 1290 Versoix, Switzerland
\and
Department of Physics and Astronomy, University of California, Los Angeles, 430 Portola Plaza, Los Angeles, CA 90095, USA
\and
School of Physics and Astronomy, University of Minnesota, 116 Church Street SE, Minneapolis, MN 55455, USA
\and
Cosmic Dawn Center (DAWN), Denmark
\and
Niels Bohr Institute, University of Copenhagen, Jagtvej 128, DK-2200 Copenhagen N, Denmark
\and
Space Telescope Science Institute, 3700 San Martin Drive, Baltimore MD, 21218
\and
Center for Astrophysical Sciences, Department of Physics and Astronomy, Johns Hopkins University, Baltimore, MD, 21218
\and
Infrared Processing and Analysis Center, Caltech, 1200 E. California Blvd., Pasadena, CA 91125, USA
\and
Centre for Astrophysics and Supercomputing, Swinburne University of Technology, PO Box 218, Hawthorn, VIC 3122, Australia
\and
School of Astronomy and Space Science, University of Chinese Academy of Sciences (UCAS), Beijing 100049, China
\and
National Astronomical Observatories, Chinese Academy of Sciences, Beijing 100101, China
\and
Institute for Frontiers in Astronomy and Astrophysics, Beijing Normal University, Beijing 102206, China
\and
School of Physics, University of Melbourne, Parkville 3010, VIC, Australia
\and
ARC Centre of Excellence for All Sky Astrophysics in 3 Dimensions (ASTRO 3D), Australia
\and
University of Ljubljana, FMF, Department of Mathematics and Physics, Jadranska ulica 19, SI-1000 Ljubljana, Slovenia
\and
Department of Physics and Astronomy, University of California Davis, 1 Shields Avenue, Davis, CA 95616, USA
\and
INAF- Osservatorio Astronomico di Padova, Vicolo Osservatorio 5, 35122 Padova, Italy
\and
Laboratory for Multiwavelength Astrophysics, School of Physics and Astronomy, Rochester Institute of Technology, 84 Lomb Memorial Drive, Rochester, NY, 14623, USA
\and 
Centre for Astrophysics and Supercomputing, Swinburne University of Technology, PO Box 218, Hawthorn, VIC 3122, Australia
}
\date{Accepted XXX. Received YYY; in original form ZZZ}

\abstract{We release fully reduced spectra obtained with NIRSpec onboard JWST as part of the GLASS-JWST Early Release Science Program and a follow-up Director's Discretionary Time program 2756. From these 263 spectra of 245 unique sources, acquired with low ($R =30-300$) and high dispersion ($R\sim2700$) gratings, we derive redshifts for 200 unique sources in the redshift range $z=0-10$. We describe the sample selection and characterize its high completeness as a function of redshift and apparent magnitude. Comparison with independent estimates based on different methods and instruments shows that the redshifts are accurate, with 80\% differing less than 0.005. We stack the GLASS-JWST spectra to produce the first high-resolution ($R \sim 2700$) JWST spectral template extending in the rest frame wavelength from 2000~\AA\ to 20, 000~\AA. Catalogs, reduced spectra, and template are made publicly available to the community.}
\maketitle

\section{Introduction}\label{sec:intro}

The commissioning of the Near InfraRed Spectrograph \citep[NIRSpec, ][]{Jakobsen2022} aboard the \textit{James Webb Space Telescope} \citep[JWST, ][]{Gardner2023} heralds a new era in our quest to explore the cosmos at near-infrared (NIR) wavelengths. With its operational range spanning from 0.6 to 5.3 microns, NIRSpec is redefining our understanding of the Universe's origins, evolution, and composition.
NIRSpec's capabilities are as diverse as they are powerful, offering astronomers the ability to capture spectra across a wide range of wavelengths with exceptional resolution and sensitivity. Equipped with three distinct spectral resolutions -- a low-dispersion prism ($R = 30-300$) and medium- ($R \sim 1000$) and high-resolution ($R \sim 2700$) gratings -- NIRSpec provides an unprecedented opportunity to study celestial objects with unparalleled precision.

Central to NIRSpec's groundbreaking capabilities is its innovative Micro-Shutter Assembly (MSA), which enables multi-object spectroscopy of many tens of objects simultaneously within a field of view of $3.6' \times 3.4'$ \citep{Ferruit2022}. This revolutionary feature empowers astronomers to conduct surveys with incomparable efficiency, assembling statistical samples of sources.

Recent observations with NIRSpec have already demonstrated its transformative potential, first by securing the spectroscopic redshift of extremely distant sources \citep[e.g., ][]{Curtis-Lake2023, Bunker2023, Roberts-Borsani2023, Roberts-Borsani2024, ArrabalHaro2023a, ArrabalHaro2023, Williams2023, Castellano2024, Wang2023, Carniani2024} either by the presence of previously inaccessible optical rest frame emission \citep[e.g., ][]{Bunker2023, Williams2023, ArrabalHaro2023, Zavala2024} or by the solid detection of the Lyman break in the continuum \citep[e.g., ][]{Curtis-Lake2023, Roberts-Borsani2023, ArrabalHaro2023a}. Additionally, the ability to detect multiple emission lines from these distant galaxies has provided invaluable insights into their physical properties and evolutionary histories by allowing us to determine the conditions of the interstellar medium such as metallicity \citep[e.g., ][]{Sanders2023, Harikane2023, Stiavelli2024, Jones2023, Curti2023, Nakajima2023}, electron temperature \citep[e.g., ][]{Tang2023, Laseter2024}, ionization parameters \citep[e.g., ][]{Cameron2023, Reddy2023}, and density \citep[e.g., ][]{Isobe2023, Boyett2023}. Moreover, NIRSpec observations have been crucial in distinguishing between active galactic nuclei (AGN) and star formation activities within galaxies \citep[e.g., ][]{Bunker2023, Castellano2024, Chisholm2024}, and in constraining the AGN contribution to galaxy evolution \citep[e.g., ][]{Maiolino2024, Larson2023}. The role of NIRSpec in studying the re-ionization era has also been significant, providing new insights into the sources and processes driving cosmic re-ionization \citep[e.g., ][]{Witstok2024, Saxena2024, Napolitano2024}.

The JWST Early Release Science Program (ERS), exemplified by the GLASS-JWST ERS program \citep[PID 1324, PI Treu;][]{Treu2022}, was designed to expedite scientific progress by providing early access to datasets that enable groundbreaking investigations. Leveraging the power of NIRSpec in combination with the other JWST instruments NIRCam and NIRISS, the GLASS-JWST ERS program has obtained deep observations of galaxies in the Hubble Frontier Field (HFF) cluster, Abell 2744 (A2744). By combining JWST's capabilities with gravitational lensing magnification \citep[][]{Bergamini2023}, this program produced breakthrough discoveries across a broad spectrum of extragalactic and high-redshift astronomical research. These include identifying high-redshift sources \citep[e.g., ][]{Castellano2022, Roberts-Borsani2022, Castellano2023, Morishita2023}, examining their morphology \citep[e.g., ][]{Yang2022, Treu2023}, ionizing properties \citep[e.g., ][]{Boyett2022, Jones2023, Mascia2023, Prieto-Lyon2023, Roy2023, Nanayakkara2023}, and exploring the general properties \citep[e.g., ][]{Wang2022, Leethochawalit2023, Santini2023, Glazebrook2023, Dressler2023} of different galaxy populations across various redshift ranges. Additionally, it has facilitated the characterization of intermediate- and low-redshift sources \citep[e.g., ][]{Vanzella2022, Marchesini2023, Jacobs2023, Vulcani2023, Chen2022, Nonino2023}, and the discovery of distant supernovae \citep{Chen2022}.

To complement the GLASS-JWST ERS program, the JWST Director's Discretionary Time (DDT) program \citep[PID 2756, PI. W. Chen;][]{Roberts-Borsani2023} on A2744 also contributed to the comprehensive exploration of the early Universe's properties and evolution.

In this work, we present the release of 263 spectra from the GLASS-JWST and the JWST-DD-2756 programs. This dataset represents a treasure trove of information, offering unprecedented insights into the spectral characteristics of high-redshift sources and their evolution over cosmic time. 

The paper is organized as follows: in Sec.~\ref{sec:data}, we characterize the sample selection, acquisition, reduction, and redshift estimation; in Sec.~\ref{sec:final}, we present the general properties of the sources; in Sec.~\ref{sec:stack}, we show the first JWST high-resolution spectral template of high-redshift galaxies and its general characteristics. Finally, in Sec.~\ref{sec:conclusions}, we summarize our key conclusions. Throughout this work, we assume a flat $\Lambda$ cold dark matter cosmology with $H_0$ = 67.7 km s$^{-1}$ Mpc$^{-1}$ and $\Omega_m$ = 0.307 \citep{Planck2020} and the \cite{Chabrier2003} initial mass function. All magnitudes are expressed in the AB system \citep{Oke1983}.

\section{Data}\label{sec:data}

\subsection{Sample selection}
NIRSpec target selection for the GLASS-JWST ERS program has been described in detail by \cite{Treu2022}. Here, we provide an updated account of the observed sources within various redshift ranges and populations outlined in the aforementioned survey paper.
 We managed to place in the MSA the following targets:
 \begin{itemize}
 \item $z>5$ spectroscopically confirmed galaxies (6) 
 \item $z>5$ extremely high Spitzer/IRAC-inferred EW(H$\alpha$) and EW([O III]) galaxies (2) 
 \item $1<z<2$ spectroscopically selected emission line galaxies from GLASS (17) 
 \item $1<z<5$ spectroscopically confirmed galaxies (10) 
 \item $z>6$ photometrically selected galaxies, including a $z\sim8$ protocluster (7) 
 \item $5<z<6$ photometrically selected galaxies (4) 
 \item $3<z<5$ photometrically selected galaxies inside the HFF WFC3 footprint (21)
 \item $z>3$ photometrically selected galaxies outside the HFF WFC3 footprint (13) 
 \item $1.7<z<4$ expected continuum sources from HFF imaging and GLASS spectroscopy (2)
 \item 53 photometrically selected sources at lower redshift, as well as those located outside of the WFC3-IR coverage but within the ACS footprint, as fillers
 \item four compact sources or alignment stars
 \item 13 serendipitously selected sources (BCG galaxies, candidate star clusters, passive galaxies, and supernovae).
 \end{itemize}

In total, our observations within the GLASS-JWST program encompassed a diverse sample of 152 sources across a range of redshifts and a variety of galaxy populations. NIRSpec target selection for the DDT program was based on NIRCam photometric calalogs, prioritizing high-redshift galaxy candidates as ``fillers'' to the supernova spectrum \citep{Roberts-Borsani2023}.

Following the selection for the GLASS-JWST ERS MSA configuration, we selected 111 sources in total, chosen to be:
\begin{itemize}
\item $z>5$ spectroscopically confirmed galaxies (4)
\item $z>5$ extremely high Spitzer/IRAC-inferred EW(H$\alpha$) and EW([O III]) galaxies (1)
\item $1<z<2$ spectroscopically selected emission line galaxies from GLASS (1)
\item $1<z<5$ spectroscopically confirmed galaxies (12)
\item $z>6$ photometrically selected galaxies, including a $z\sim8$ protocluster (5)
\item $5<z<6$ photometrically selected galaxies (3)
\item $3<z<5$ photometrically selected galaxies (18)
\item $z>3$ photometrically selected galaxies outside the HFF WFC3 footprint (32)
\item four compact sources or alignment stars
\item 31 peculiar objects at different redshifts
\item 31 serendipitously selected sources.
\end{itemize}

\begin{figure*}[ht!]
\centering
\includegraphics[width=1\linewidth]{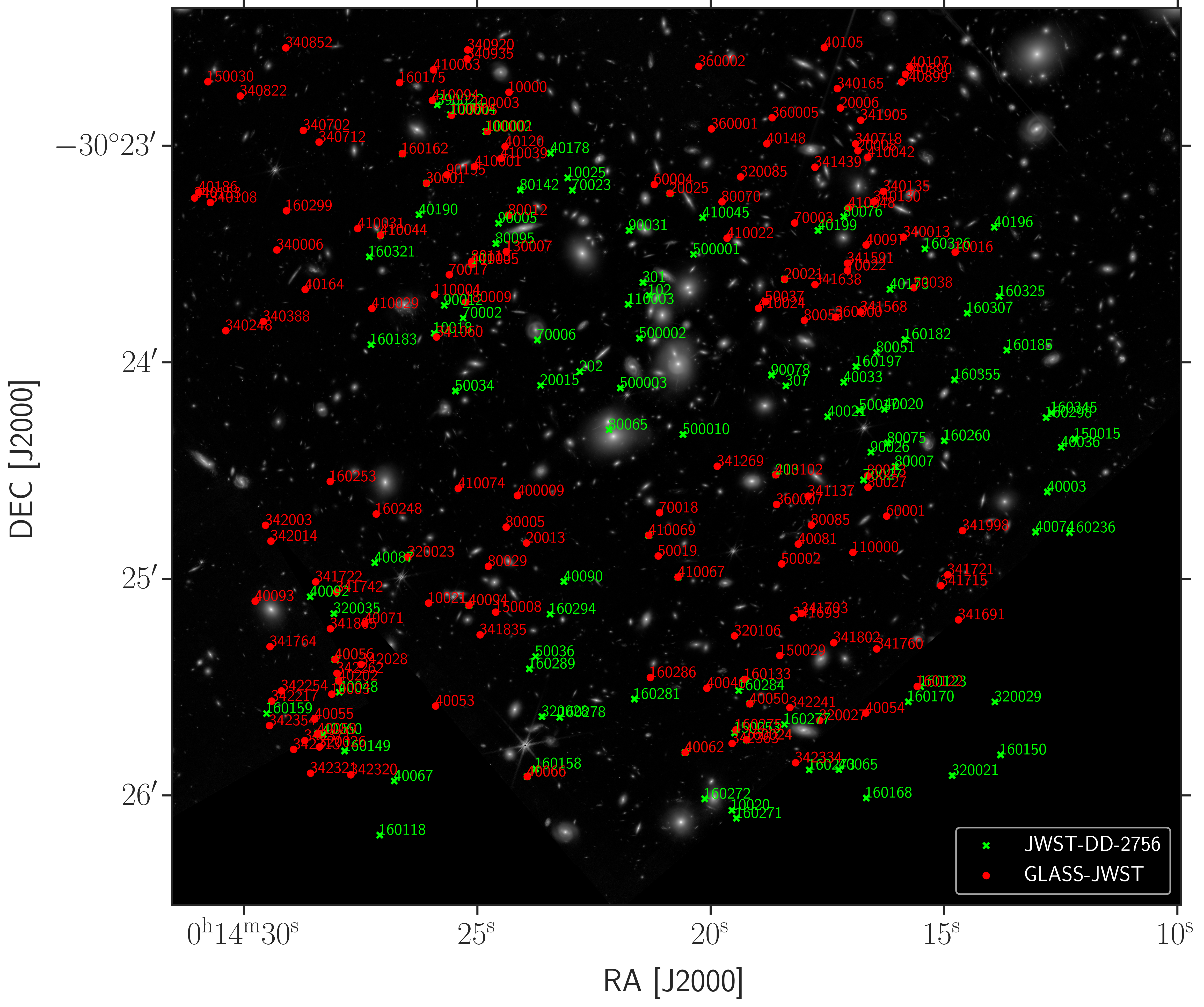}
\caption{Spatial distribution of the 245 observed sources overlaid on the UNCOVER program (PID 2561, PI I. Labbé) image obtained with the NIRCam filter F200W. 
The observed sources are color-coded based on their program: GLASS-JWST ERS sources are marked in red, and JWST DDT sources (PID 2756) in green.}\label{fig:loc}
\end{figure*}

The spatial distribution of the selected sources, color-coded by their respective programs, can be visualized in Fig. \ref{fig:loc}.

\subsection{Data acquisition}

Spectra were acquired through NIRSpec MSA observations in two distinct programs: the GLASS-JWST ERS program and the JWST DDT program PID 2756.
In the GLASS-JWST observations, conducted on November 10, 2022, three spectral configurations (G140H/F100LP, G235H/F170LP, and G395H/F290LP) were employed, covering a wavelength range from 1 to 5.14 $\mu$m with a spectral resolution of approximately 2000-3000 \citep{Jakobsen2022}. 
An NRSIRS2 readout pattern was adopted.  Each NIRSpec band was given 20 Groups, one integration, and two exposures.  The choice was in part motivated by the need to observe NIRCam in parallel with seven bands \citep{Treu2022}. Each dither or nod point takes 1473.5 seconds; with three nods, two sub-shutter dithers, and two separate exposures, the total time per band is 1473.5 seconds times 12, or 4.9 hours.

On October 23, 2022, the DDT NIRSpec observations were conducted using the CLEAR filter+PRISM configuration, which provided continuous wavelength coverage from 0.6 to 5.3 $\mu$m at a spectral resolution of approximately 30-300 \citep{Jakobsen2022}. An NRS readout pattern was adopted, with 17 Groups and one integration. Each dither or nod point takes 741 seconds; with three nods, and two separate exposures, the total exposure time per MSA configuration is 2222s. Some objects are observed in both MSA configurations, resulting in an on-source exposure time for these observations of 1.23 hours.

\subsection{Data reduction}

The data reduction was carried out using the official STScI JWST pipeline (ver.1.8.2)\footnote{\url{https://github.com/spacetelescope/jwst}} for Level 1 data products, and the \texttt{msaexp}\footnote{\url{https://github.com/gbrammer/msaexp}} code for Level 2 and 3 data products, which is based on the STScI pipeline but also includes additional correction routines. In summary, we initially reduced the uncalibrated data using the \texttt{Detector1Pipeline} routine and the latest set of reference files (jwst\_1023.pmap) to correct for detector-level artifacts and convert them to count-rate images. Then, we applied custom preprocessing routines from \texttt{msaexp} to remove residual $1/f$ noise that is not corrected by the IRS2 readout, identify and remove snowballs, and remove bias exposure by exposure before running STScI routines from \texttt{Spec2Pipeline} for the final 2D cutout images. To perform WCS registration, flat-fielding, path-loss corrections, and flux calibration, these routines include \texttt{AssignWcs}, \texttt{Extract2dStep}, \texttt{FlatFieldStep}, \texttt{PathLossStep}, and \texttt{PhotomStep}. Of note, our chosen reference files include an in-flight flux calibration, accounting for NIRSpec's better-than-expected throughput at blue wavelengths. Local background subtraction was performed using a three-shutter nod pattern before the resulting images were drizzled onto a common grid. We optimally extracted the spectra using an inverse-variance weighted kernel, which was derived by summing the 2D spectrum along the dispersion axis and fitting the signal along the spatial axis to a Gaussian profile. We visually inspected all kernels to make sure spurious events were not included. As a result, the kernel extracts the 1D spectrum along the dispersion axis. The final step was to verify the default wavelength calibration for the gratings, which is accurate within 1 \AA. 

\begin{figure*}[ht!]
\includegraphics[width=1\linewidth]{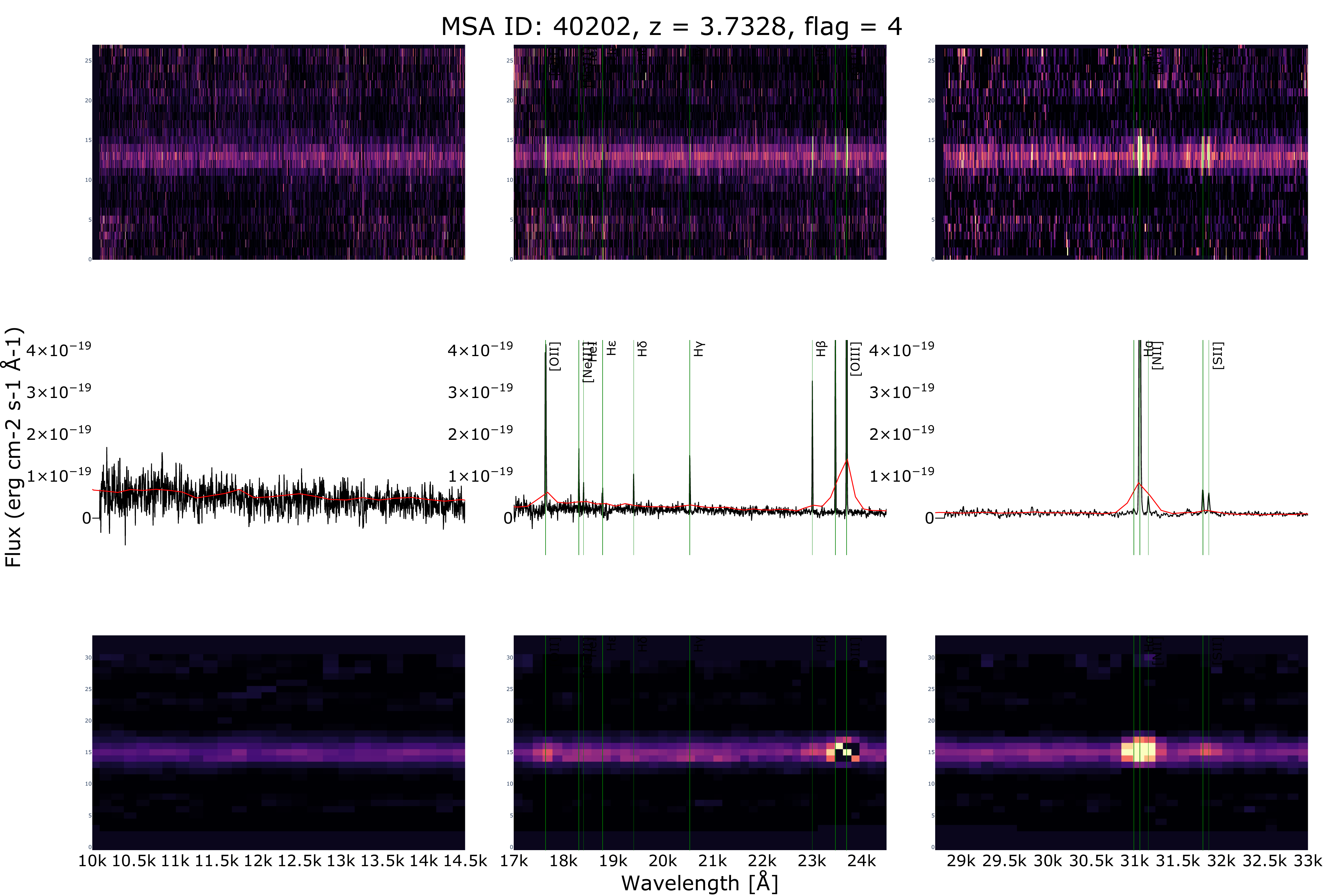}
\caption{Example of GLASS-JWST and JWST-DD-2756 spectra of the same source (MSA ID: 40202). Upper panel: 2D spectrum obtained with the GLASS-JWST program ($R \sim 2700$). Middle panel: 1D extracted spectra for both the GLASS-JWST program (in black) and the JWST-DD-2756 program (in red). Lower panel: 2D spectrum obtained with the JWST-DD-2756 program ($R \sim 100$).\label{fig:full_data}}
\end{figure*}

Examples of the final products from the data reduction process are shown in Fig.~\ref{fig:full_data}, in which we present spectra of the same source observed with both programs. As can be observed, the prism can detect lines in addition to the continuum; however, in order to spectrally resolve and split doublets, higher resolutions are needed, such as those achieved with the GLASS-JWST program.

\subsection{Redshift estimation, reliability flags, and confidence levels}\label{sec:lime}

In most cases, the redshift of each spectrum was determined through emission line measurements, without any prior knowledge of photometric redshifts. For sources displaying multiple distinct emission lines, identifying the [O III] and/or H$\alpha$ emission lines was straightforward. Spectroscopic redshifts were determined by comparing the observed emission lines with their corresponding theoretical vacuum transitions. To establish the transition wavelength for each spectral line, we applied a Gaussian profile fitting. In the case of the H$\beta$ + [O III] emission lines, we anticipated that they would all share the systemic redshift. Consequently, we performed a collective fit for these lines and utilized the Gaussian centroids to calculate the mean redshifts. For objects featuring only a single emission line, we employed spectral line fitting to determine the transition wavelength and estimate a possible redshift. For this part of our analysis, we used \textsc{Mpfit}\footnote{\url{http://purl.com/net/mpfit}} \citep{markwardt2009}.

For all other sources, for which no obvious emission lines were apparent, we fit for the redshift of the sources using \texttt{msaexp} and assuming their photometric redshift with an uncertainty of $\Delta z\pm0.1$.

 The redshift values determined using this approach underwent independent verification. Several authors conducted visual inspections of the spectra, either for all of them or for a subsample, and consistently obtained the same results.

Redshift measurements are categorized as follows:
\begin{itemize}
\item Flag 4: Represents a secure redshift, with an estimated accuracy of above 99\%, based on multiple spectral emission features (119).
\item Flag 3: Denotes a very reliable redshift, comparable in confidence to Flag 4, and supported by a single clear spectral emission feature (15).
\item Flag 2: Signifies a fairly reliable redshift, with confidence level just below that of Flags 3 and 4. It is supported by cross-correlation results and photometric redshift data (8).
\item Flag 1: Indicates no reliable spectroscopic redshift measurement due to the lack of any emission line (51).
\item Flag 14: Defines a spectrum with no emission lines; the redshift estimate is based on fitting the continuum (70).
\end{itemize}

\begin{figure*}[!ht]
\centering
\includegraphics[width=0.95\linewidth]{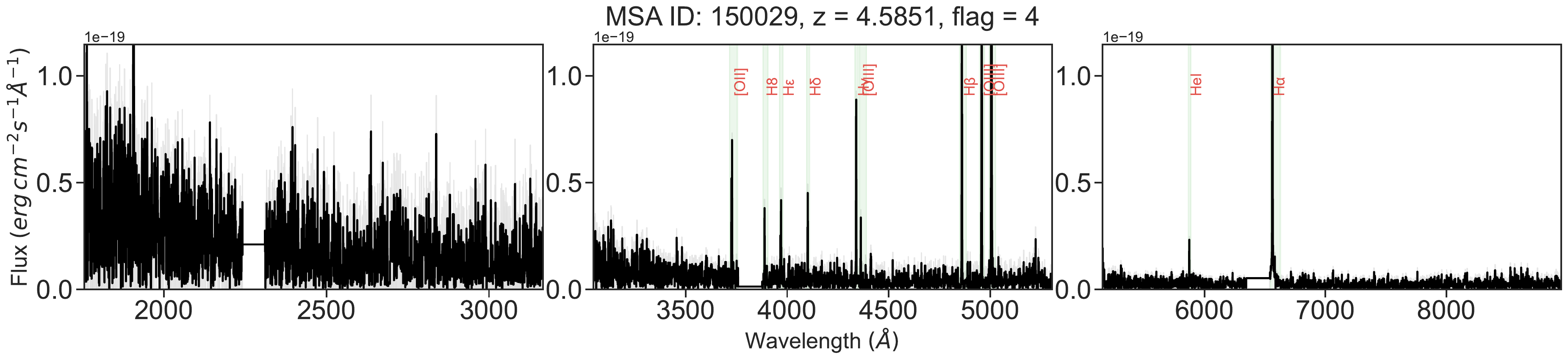}\\
\includegraphics[width=0.95\linewidth]{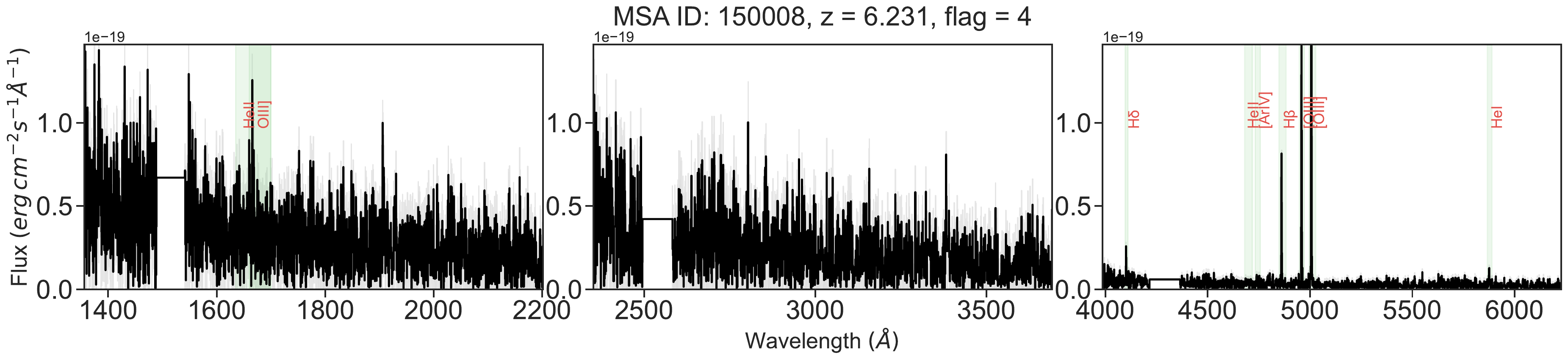}\\
\includegraphics[trim=-20 0 10 0, clip, width=0.95\linewidth]{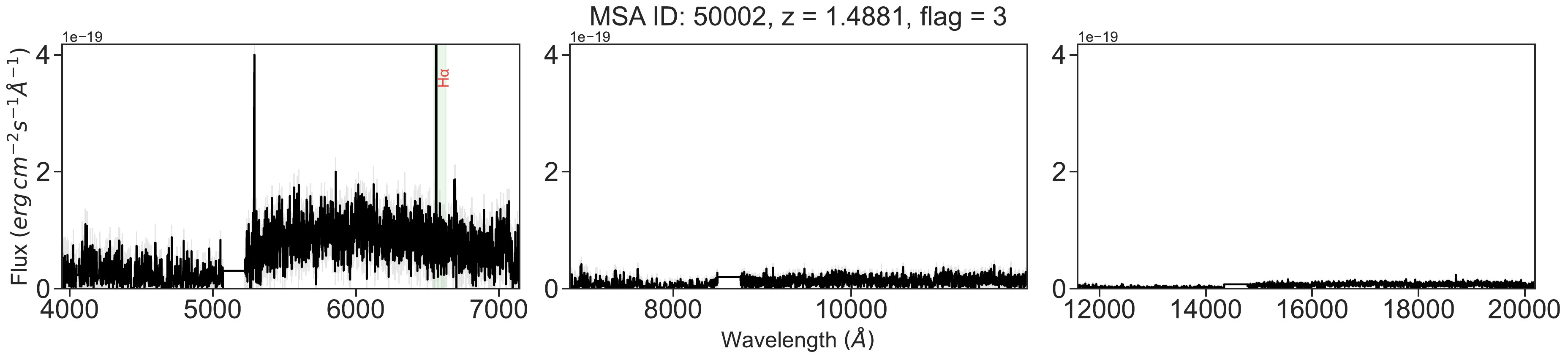}\\
\includegraphics[trim=-20 0 5 0, clip, width=0.95\linewidth]{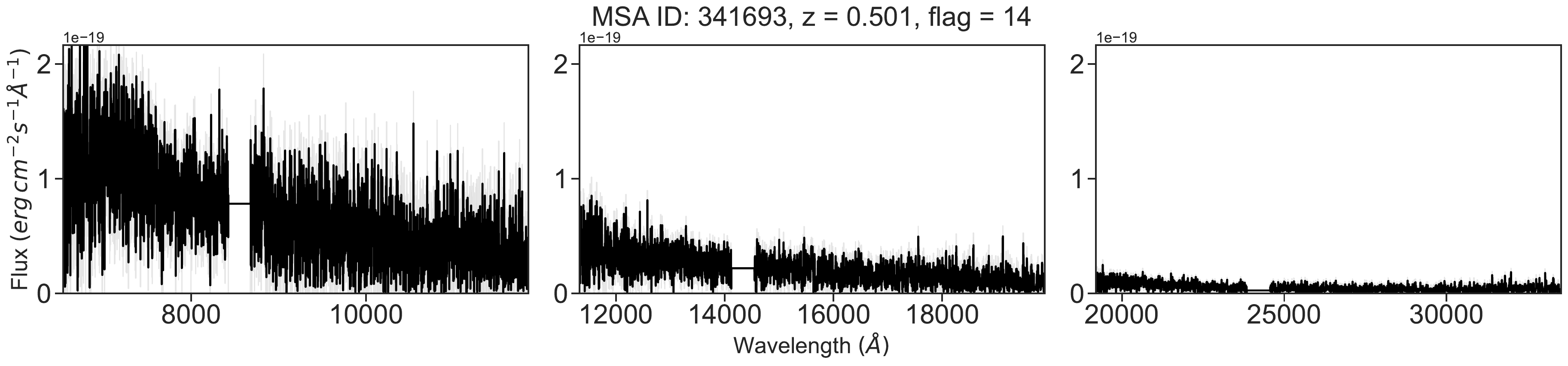}\\
\caption{Examples of GLASS-JWST spectra at different redshifts. A zoom into the spectra is shown around the region containing the most prominent lines, according to the galaxy redshift. The redshift and reliability flag for each galaxy are indicated in each panel.} \label{fig:GLASS_spectra}
\end{figure*}

\begin{figure*}[!ht]
\centering
\includegraphics[width=0.9\linewidth]{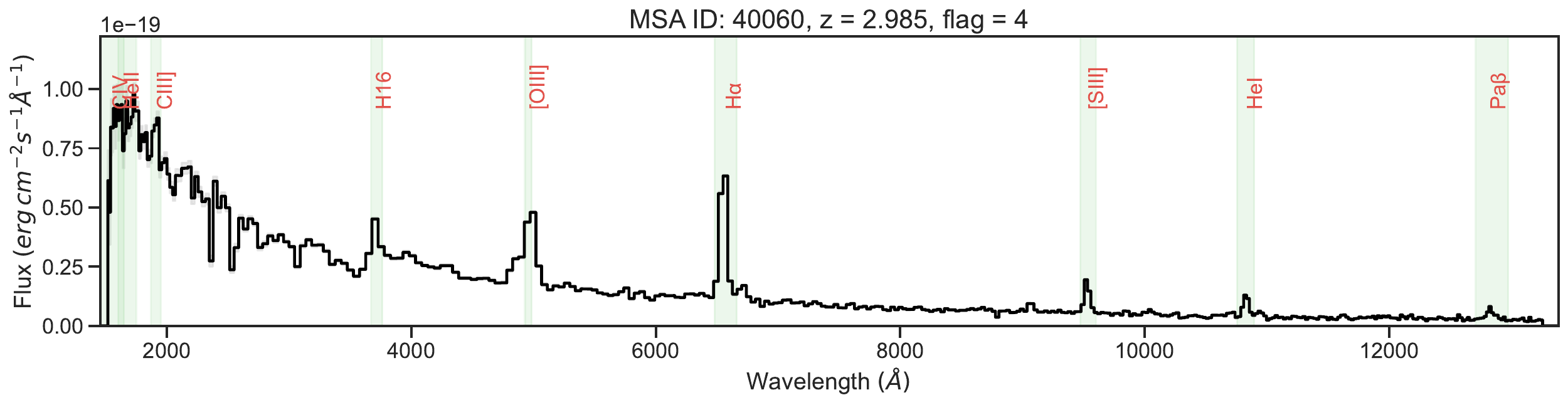}\\
\includegraphics[width=0.92\linewidth]{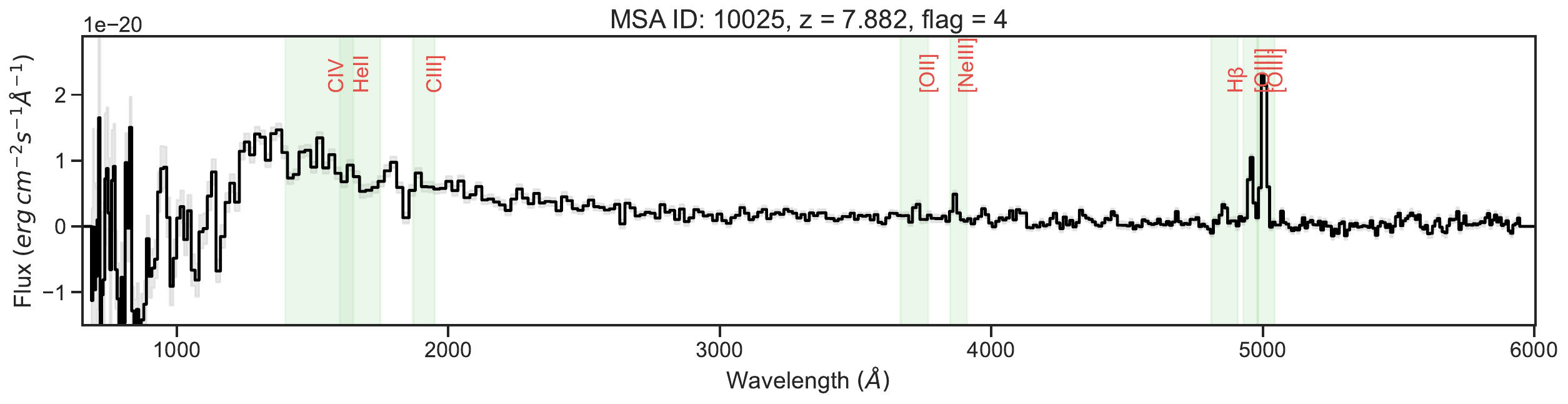}\\
\includegraphics[width=0.9\linewidth]{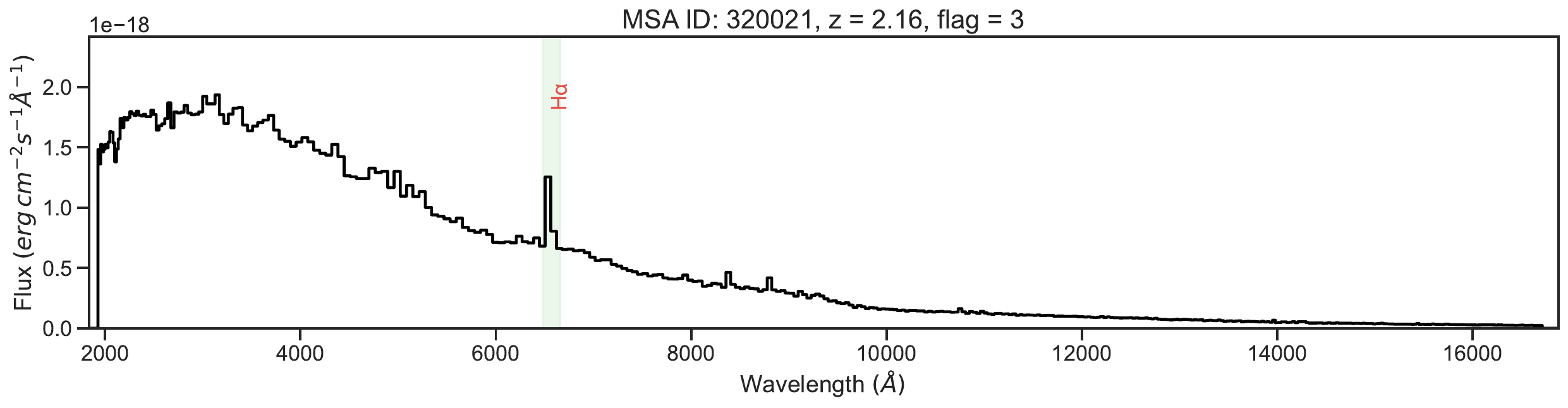}\\
\includegraphics[width=0.9\linewidth]{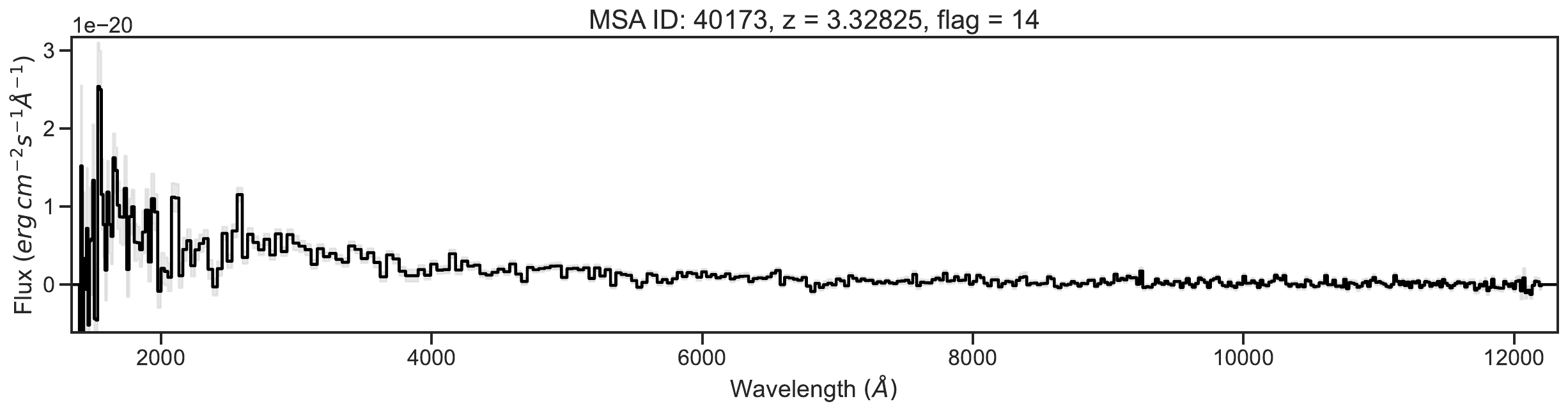}
\caption{Examples of JWST-DD-2756 spectra at different redshifts. The redshift and reliability flag for each galaxy are indicated in each panel.\label{fig:DDT_spectra}} 
\end{figure*}

We utilized LiMe\footnote{\url{https://lime-stable.readthedocs.io/en/latest/}} \citep{Fernandez2024} to compile the line catalog for each source using the spectroscopic redshift information. Ensuring reliability, we established a minimum signal-to-noise ratio (S/N) threshold of five for line identification, assuming a Gaussian profile. This approach was applied to both the DDT and GLASS-JWST spectra, and illustrative examples are presented in Fig.~\ref{fig:GLASS_spectra} and~\ref{fig:DDT_spectra}. The complete list of all detected emission lines across all spectra is provided in Table A.\ref{tab:spectral_lines}.

\subsection{Comparison with independent measurements}

Among the GLASS-JWST + JWST-DD-2756 targets with available redshift measurements, 96 (with repetitions) objects previously had their redshifts published in the literature \citep[e.g., ][]{Roberts-Borsani2023, Morishita2023, Prieto-Lyon2023, Mascia2023, Vulcani2023, Roy2023, Nakajima2023}. These published redshift measurements were derived from the same publicly available data, albeit in some cases reduced based on earlier versions of the pipeline and calibration files. When comparing with published values, one needs to take into account these differences, as well differences in measurement technique. For example, our redshift estimation relies on emission line measurements, whereas others use template fitting, which may result in minor discrepancies. 

Moreover, some of the objects were also observed with other instruments, such as VLT-MUSE \citep{Richard2021}. 

\begin{figure}[ht!]
\includegraphics[width=8cm]{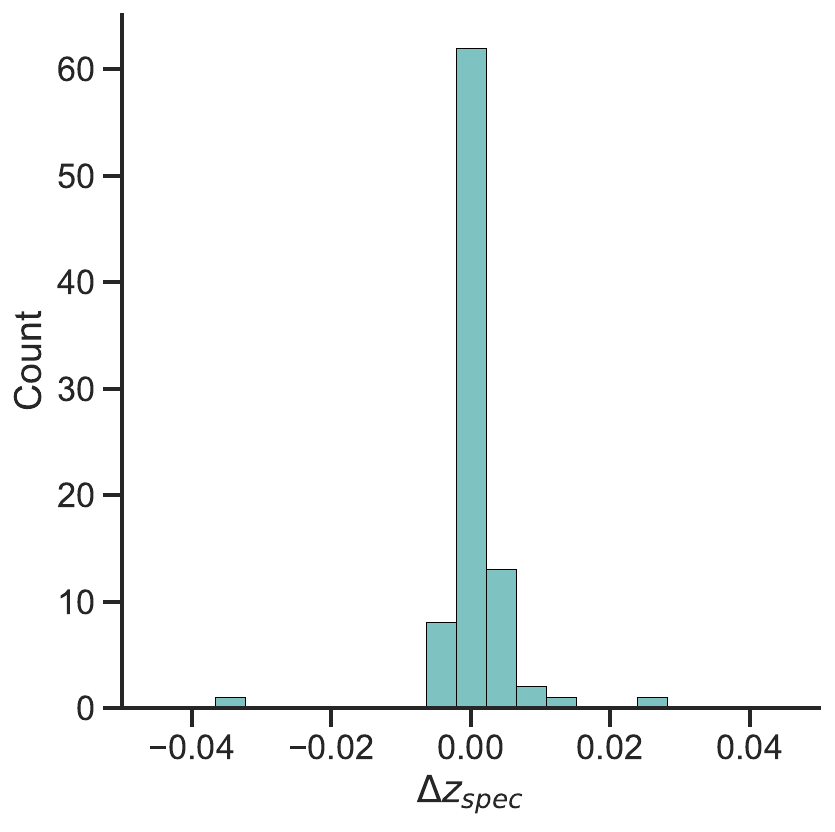}
\caption{Comparison of redshift measurements between our study and previously published values for the GLASS-JWST + JWST-DD-2756 targets.\label{fig:z_comp}}
\end{figure}

We compared all published values to our measurements, regardless of their quality.The results are shown in Fig. \ref{fig:z_comp}. The distribution exhibits a sharp peak at zero, with approximately 80\% of the measurements differing by less than 0.005. We cut the x axis at $\pm 0.05$, noting that there are only three sources outside this range. These outliers were selected from the \cite{Richard2021} catalog and are flagged as 14 (no emission lines) or 3 (one single clear spectral emission feature).

\section{The final sample} \label{sec:final}
There is a small overlap between the JWST-DD-2756 and the GLASS-JWST catalogs, with a total of 18 sources appearing in both datasets. Among these dual-listed sources, a detailed examination reveals that 12 sources have reliable redshift estimates in both low- and high-resolution spectra. Three of the sources have a reliable redshift estimate exclusively in the GLASS-JWST data,
two sources get a redshift solution solely in the JWST-DD-2756 data, and one source did not yield any viable solution in either catalog.

This analysis underscores the subtleties of redshift determination and highlights the importance of cross-referencing multiple datasets to obtain comprehensive insights into the properties of astronomical objects.

\begin{figure}[ht!]
\includegraphics[width=8cm]{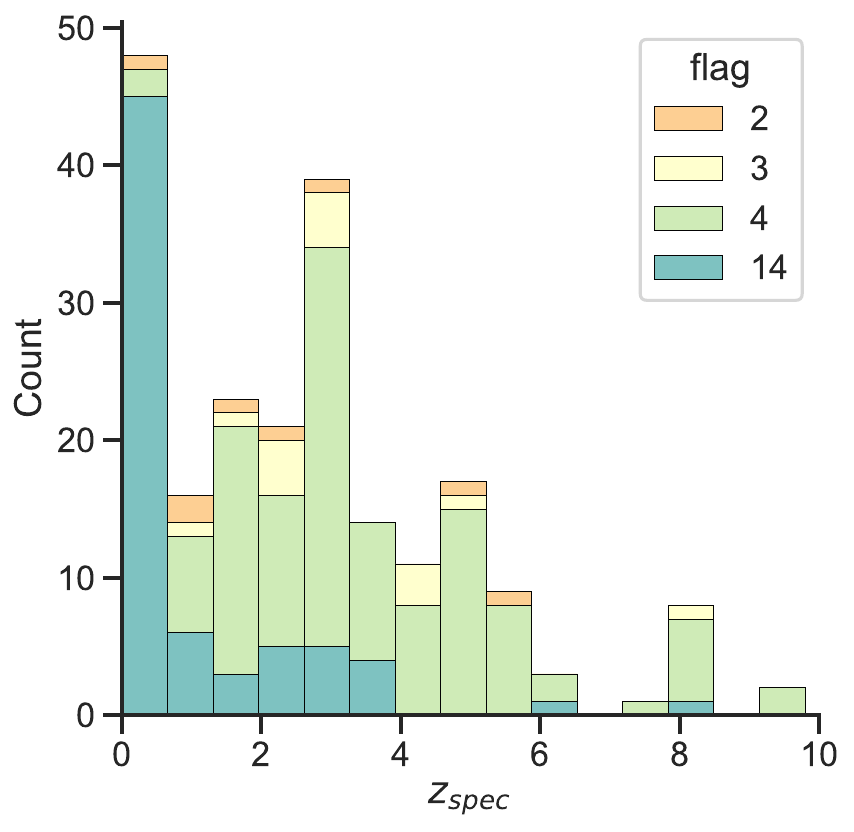}
\caption{Redshift distribution of the final GLASS-JWST + JWST-DD-2756 sample. \label{fig:z_distribution}}
\end{figure}

Figure \ref{fig:z_distribution} shows the redshift distribution of the final GLASS-JWST + JWST-DD-2756 spectroscopic sample.
In orange, yellow, green, and blue, the distributions for flags 2, 3, 4, and 14 are shown, respectively. Overall, we have determined the redshift for 200 unique sources, with a median redshift of $z=2.61$.

Fifty-one spectra lack redshift confirmation and were flagged as 1. Of these targets, 37 were initially chosen with a photometric redshift higher than 3, implying detectability through emission lines. However, in the updated version of the photometric catalog (Merlin et al. in prep), their redshifts are now below 0.5. For these redshifts, we cannot detect the breaks as they fall outside of the observed range, and sources are expected to have modest emission lines, especially if they are not actively star-forming. Therefore, the fact that we detect only featureless continua is consistent with their low-redshift nature, but prevents us from determining spectroscopic redshifts. 
In addition, eight sources were initially selected from the MUSE catalog within the redshift range from 3 to 6. Among these, four have a confidence level of 3 according to the \cite{Richard2021} catalog, suggesting redshift determination based on multiple spectral features or additional information on a high S/N emission line, while the remaining four have a confidence level of 2, indicating that their redshifts are derived from a single emission line without supplementary data.

Out of the 51 spectra lacking redshift confirmation and flagged as 1, five sources have a redshift determination in one of the two programs. Only one source, observed in both programs, does not exhibit any spectral features. Thus, in the end, there are 45 unique sources with no redshift estimate.

\begin{figure*}[ht!]
\sidecaption
\includegraphics[width=12cm]{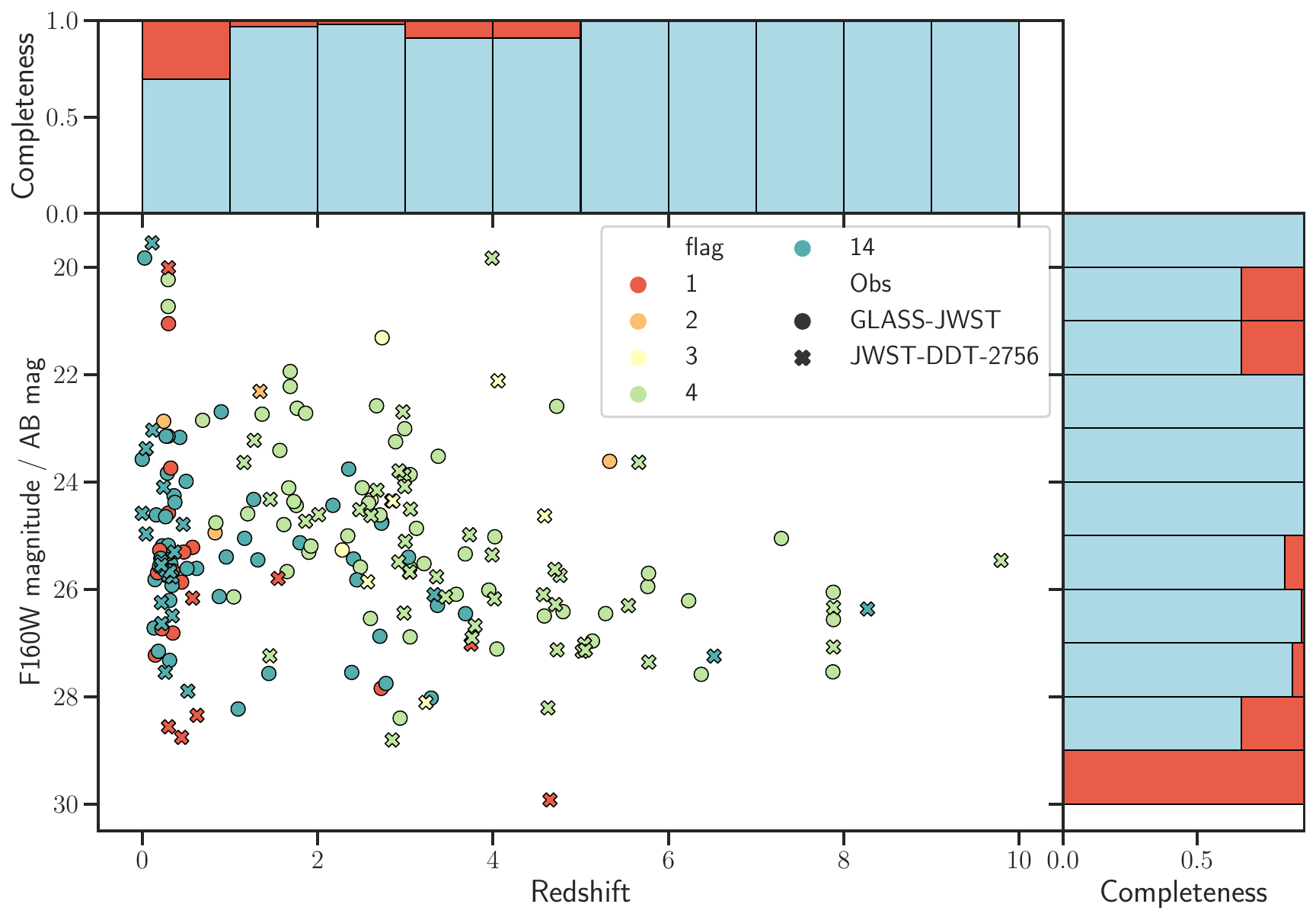}
\caption{Redshift distribution as a function of the F160W magnitude, with symbols indicating observations from the GLASS-JWST program (circles) and the JWST-DDT-2756 program (x marks). The color of each data point corresponds to its respective flag value \citep[follow-up of Fig. 6 in ][]{Treu2022}. In the upper panel, we display the completeness in redshift, with red indicating the percentage of sources with no redshift estimation, while in light blue, we represent the total fraction normalized to 1 of the sources in the bin. Similarly, in the right panel, we show the completeness in the F160W magnitude, following the same color scheme.\label{fig:mag_z}}
\end{figure*}

In Fig.~\ref{fig:mag_z}, we illustrate the redshift distribution relative to the F160W magnitude for both programs, alongside the completeness in redshift and the F160W magnitude. Notably, a significant portion of sources lacking redshift determination are photometrically identified as low-redshift sources; hence, the absence of spectral features. 
Conversely, we do not observe any discernible trend between completeness and the F160W magnitude, even though, as was expected, the faintest sources tend to lack redshift determination. 

\section{Spectral stacking} \label{sec:stack}

In order to produce the first high-resolution JWST spectral template of high-redshift galaxies, we stacked all the sources from the GLASS-JWST program with spectra flagged as quality 3 or 4. In total, we used 74 sources, spanning redshifts from 0.3 to 9.3. The median redshift of the stacked spectra is 2.813. 

\begin{figure*}[ht!]
\centering
\includegraphics[origin=c, width=\linewidth]{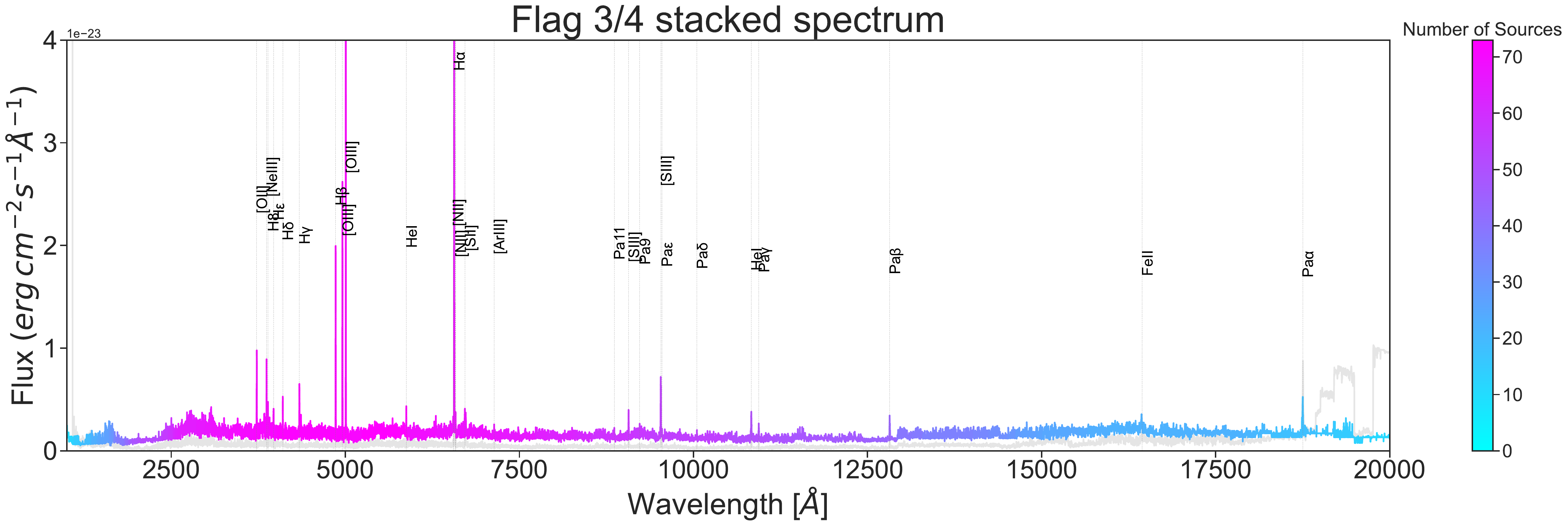}
\includegraphics[origin=c, width=\linewidth]{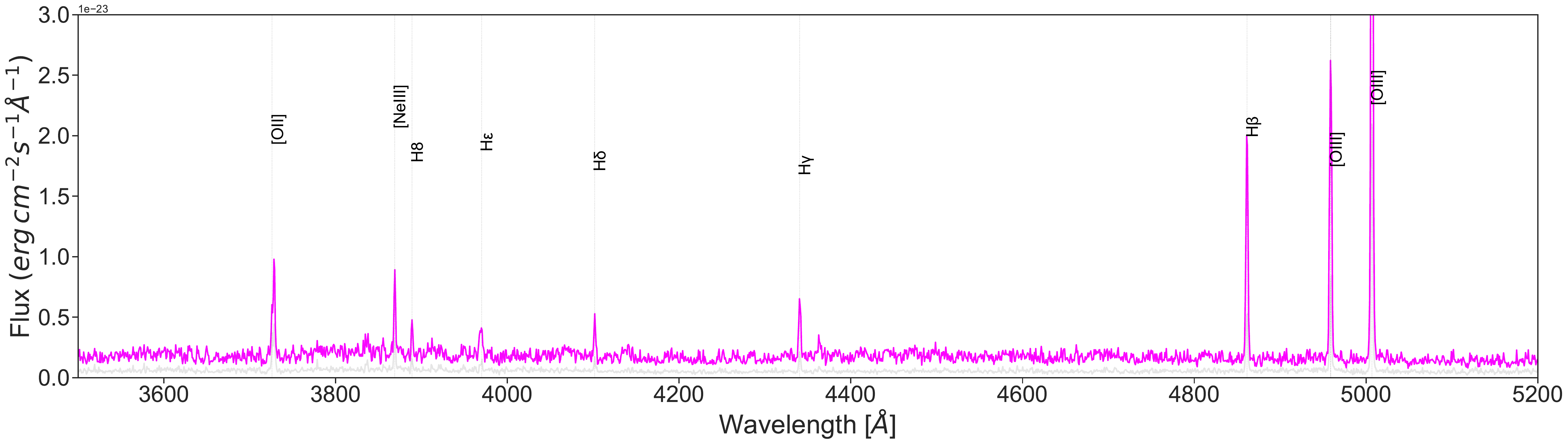}
\includegraphics[origin=c, width=\linewidth]{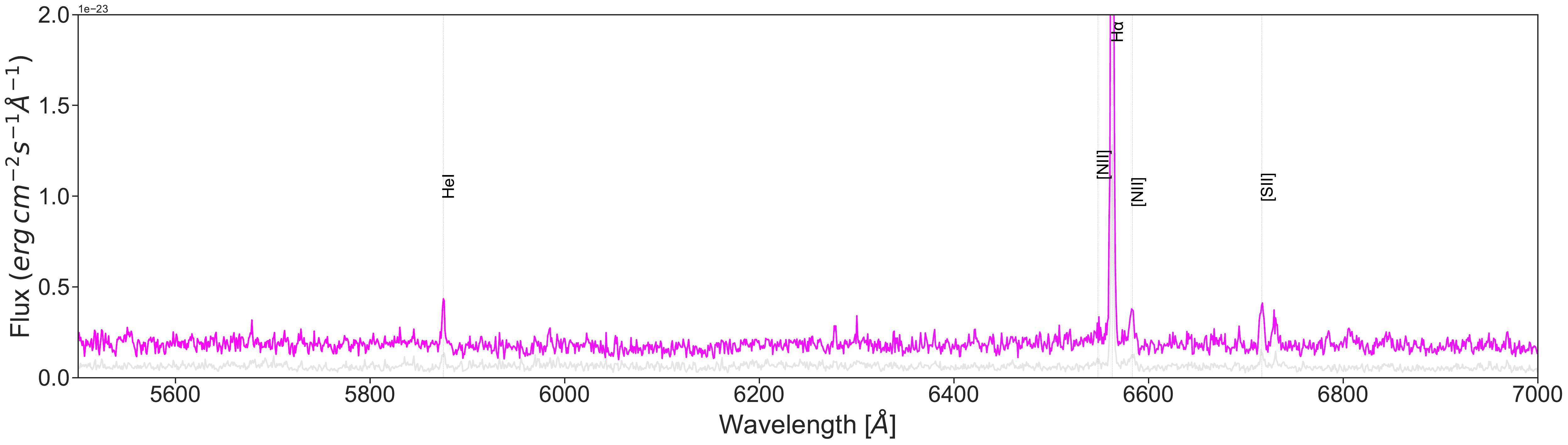}
\includegraphics[origin=c, width=\linewidth]{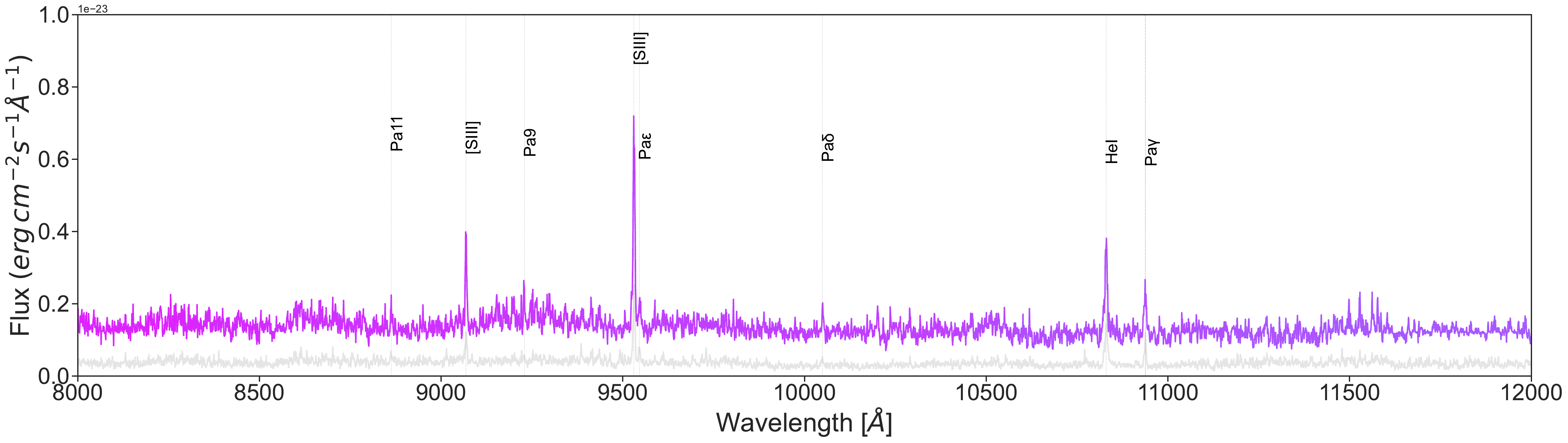}
\caption{Stacked spectrum derived from 74 sources identified with flags 3 and 4 from the GLASS-JWST program. These sources span redshifts ranging from $z \sim 0$ to $z \sim 9$. Noise is shown in grey. The flux is color-coded based on the number of contributing sources.\label{fig:stack}}
\end{figure*}

The stacking was performed by first converting each spectrum to the rest-frame, using the spectroscopic redshift. 
The rest-frame spectra were first normalized using the mean flux density value in the rest-frame wavelength range of 10000 \AA $\ <\lambda <11000$ \AA\ and the AB magnitude in the F160W filter. The spectra were then resampled to a wavelength grid ranging from 0 to 27000 \AA, with a step size of 0.6 \AA (the wavelength resolution obtained at a redshift of 3.3, which is the mean redshift of sources in this selection). The errors on the stacked spectra were calculated using bootstrapping: we randomly sampled and stacked the same number of galaxies from the sample 500 times, and the dispersion on fluxes thus obtained gives the errors (Fig. \ref{fig:stack}).

\section{Summary} \label{sec:conclusions}

This study presents the spectroscopic data release from the GLASS-JWST and JWST-DD-2756 programs conducted on the Abell2744 cluster field, derived from a comprehensive investigation with the JWST's NIRSpec employing different dispersion gratings, with low ($R =30-300$) and high dispersion ($R\sim2700$). The objective is to address fundamental questions on the high-redshift Universe such as the sources responsible for ionizing the universe and understanding the cycling of baryons through galaxies, which are the key drivers of this investigation.

We examined a total of 263 targets, successfully achieving robust redshift determinations for 206 of them spanning a range from z=0 to z=9.793, with a F160W AB magnitude range from 19 to 30. Notable discoveries include the identification of 16 galaxies at redshifts greater than 6. Approximately 80\% of the targeted galaxies have been spectroscopically confirmed within the expected redshift range. The study's success can be attributed to the meticulous process of target selection, utilizing spectrophotometric catalogs from our team and existing literature on the A2744 cluster.

For all of the galaxies, we release the full 2D and 1D spectra, the redshift and its quality flag, and the list of the detected emission lines with an S/N higher than 5.

Additionally, we present the first high-resolution stacked spectrum from JWST observations, spanning the rest-frame range from 1200 to 20000 \AA, showcasing JWST's capability to capture signals from multiple optical and NIR emission lines in star-forming galaxies.

In terms of scientific significance, the study underscores the effectiveness of NIRSpec observations in achieving the goals set forth by the GLASS-JWST ERS program. By prioritizing the observation of rarer, high-redshift galaxies while also including numerous lower-redshift galaxies, the study supports a comprehensive exploration of cosmic evolution from the epoch of cosmic noon ($z\sim2$) to the era of reionization ($z\geq5$).
The catalog, spectra, and stacked spectrum, have been made publicly available\footnote{\url{https://archive.stsci.edu/hlsp/glass-jwst}} to facilitate the exploitation of this dataset.

\begin{acknowledgements}
This work is based on observations made with the NASA/ESA/CSA \textit{James Webb Space Telescope} (JWST). The data were obtained from the Mikulski Archive for Space Telescopes at the Space Telescope Science Institute, which is operated by the Association of Universities for Research in Astronomy, Inc., under NASA contract NAS 5-03127 for JWST. These observations are associated with program JWST-ERS-1342, and JWST-DD-2756. We acknowledge financial support from NASA through grant JWST-ERS-1342. We also acknowledge support from the INAF Large Grant 2022 “Extragalactic Surveys with JWST” (PI Pentericci).
KB is supported by the Australian Research Council Centre of Excellence for All Sky Astrophysics in 3 Dimensions (ASTRO 3D), through project number CE170100013. MB acknowledges support from the ERC Grant FIRSTLIGHT and from the Slovenian national research agency ARIS through grants N1-0238 and P1-0188. BV is supported  by the European Union – NextGenerationEU RFF M4C2 1.1 PRIN 2022 project 2022ZSL4BL INSIGHT.
\end{acknowledgements}

\section*{Data availability} \label{sec:data_release}
 
The public data release, available from the Mikulski Archive for Space Telescopes\footnote{\url{https://archive.stsci.edu/hlsp/glass-jwst}}, includes the following components:

\begin{enumerate}
\item A catalog: this contains MSA ID, RA, and DEC coordinates, the program, spectroscopic redshift, accuracy flag, and a list of detected emission lines with an S/N higher than 5 for all of the 263 detected sources. The complete list of emission lines is presented in Table A.\ref{tab:spectral_lines}.
\item Reduced and calibrated 1D and 2D spectra. 
\item Spectral template of high-redshift galaxies: this is a single FITS file containing extensions for \texttt{flux}, \texttt{wave}, and \texttt{noise}, which represent flux, wavelength, and noise information computed in the manner described in Sec. \ref{sec:stack}. 
\end{enumerate}

\bibliography{sample631}{}

\begin{thebibliography}{62}
\expandafter\ifx\csname natexlab\endcsname\relax\def\natexlab#1{#1}\fi

\bibitem[{{Arrabal Haro} {et~al.}(2023{\natexlab{a}}){Arrabal Haro},
  {Dickinson}, {Finkelstein}, {Fujimoto}, {Fern{\'a}ndez}, {Kartaltepe},
  {Jung}, {Cole}, {Burgarella}, {Chworowsky}, {Hutchison}, {Morales},
  {Papovich}, {Simons}, {Amor{\'\i}n}, {Backhaus}, {Bagley}, {Bisigello},
  {Calabr{\`o}}, {Castellano}, {Cleri}, {Dav{\'e}}, {Dekel}, {Ferguson},
  {Fontana}, {Gawiser}, {Giavalisco}, {Harish}, {Hathi}, {Hirschmann},
  {Holwerda}, {Huertas-Company}, {Koekemoer}, {Larson}, {Lucas}, {Mobasher},
  {P{\'e}rez-Gonz{\'a}lez}, {Pirzkal}, {Rose}, {Santini}, {Trump}, {de la
  Vega}, {Wang}, {Weiner}, {Wilkins}, {Yang}, {Yung}, \&
  {Zavala}}]{ArrabalHaro2023a}
{Arrabal Haro}, P., {Dickinson}, M., {Finkelstein}, S.~L., {et~al.}
  2023{\natexlab{a}}, \apjl, 951, L22

\bibitem[{{Arrabal Haro} {et~al.}(2023{\natexlab{b}}){Arrabal Haro},
  {Dickinson}, {Finkelstein}, {Kartaltepe}, {Donnan}, {Burgarella}, {Carnall},
  {Cullen}, {Dunlop}, {Fern{\'a}ndez}, {Fujimoto}, {Jung}, {Krips}, {Larson},
  {Papovich}, {P{\'e}rez-Gonz{\'a}lez}, {Amor{\'\i}n}, {Bagley}, {Buat},
  {Casey}, {Chworowsky}, {Cohen}, {Ferguson}, {Giavalisco}, {Huertas-Company},
  {Hutchison}, {Kocevski}, {Koekemoer}, {Lucas}, {McLeod}, {McLure}, {Pirzkal},
  {Seill{\'e}}, {Trump}, {Weiner}, {Wilkins}, \& {Zavala}}]{ArrabalHaro2023}
{Arrabal Haro}, P., {Dickinson}, M., {Finkelstein}, S.~L., {et~al.}
  2023{\natexlab{b}}, \nat, 622, 707

\bibitem[{{Bergamini} {et~al.}(2023){Bergamini}, {Acebron}, {Grillo}, {Rosati},
  {Caminha}, {Mercurio}, {Vanzella}, {Mason}, {Treu}, {Angora}, {Brammer},
  {Meneghetti}, {Nonino}, {Boyett}, {Brada{\v{c}}}, {Castellano}, {Fontana},
  {Morishita}, {Paris}, {Prieto-Lyon}, {Roberts-Borsani}, {Roy}, {Santini},
  {Vulcani}, {Wang}, \& {Yang}}]{Bergamini2023}
{Bergamini}, P., {Acebron}, A., {Grillo}, C., {et~al.} 2023, \apj, 952, 84

\bibitem[{{Boyett} {et~al.}(2022){Boyett}, {Mascia}, {Pentericci},
  {Leethochawalit}, {Trenti}, {Brammer}, {Roberts-Borsani}, {Strait}, {Treu},
  {Bradac}, {Glazebrook}, {Acebron}, {Bergamini}, {Calabr{\`o}}, {Castellano},
  {Fontana}, {Grillo}, {Henry}, {Jones}, {Marchesini}, {Mason}, {Mercurio},
  {Morishita}, {Nanayakkara}, {Rosati}, {Scarlata}, {Vanzella}, {Vulcani},
  {Wang}, \& {Willott}}]{Boyett2022}
{Boyett}, K., {Mascia}, S., {Pentericci}, L., {et~al.} 2022, \apjl, 940, L52

\bibitem[{{Boyett} {et~al.}(2024){Boyett}, {Trenti}, {Leethochawalit},
  {Calabr{\'o}}, {Metha}, {Roberts-Borsani}, {Dalmasso}, {Yang}, {Santini},
  {Treu}, {Jones}, {Henry}, {Mason}, {Morishita}, {Nanayakkara}, {Roy}, {Wang},
  {Fontana}, {Merlin}, {Castellano}, {Paris}, {Brada{\v{c}}}, {Malkan},
  {Marchesini}, {Mascia}, {Glazebrook}, {Pentericci}, {Vanzella}, \&
  {Vulcani}}]{Boyett2023}
{Boyett}, K., {Trenti}, M., {Leethochawalit}, N., {et~al.} 2024, Nature
  Astronomy [\eprint[arXiv]{2303.00306}]

\bibitem[{{Bunker} {et~al.}(2023){Bunker}, {Saxena}, {Cameron}, {Willott},
  {Curtis-Lake}, {Jakobsen}, {Carniani}, {Smit}, {Maiolino}, {Witstok},
  {Curti}, {D'Eugenio}, {Jones}, {Ferruit}, {Arribas}, {Charlot}, {Chevallard},
  {Giardino}, {de Graaff}, {Looser}, {L{\"u}tzgendorf}, {Maseda}, {Rawle},
  {Rix}, {Del Pino}, {Alberts}, {Egami}, {Eisenstein}, {Endsley}, {Hainline},
  {Hausen}, {Johnson}, {Rieke}, {Rieke}, {Robertson}, {Shivaei}, {Stark},
  {Sun}, {Tacchella}, {Tang}, {Williams}, {Willmer}, {Baker}, {Baum},
  {Bhatawdekar}, {Bowler}, {Boyett}, {Chen}, {Circosta}, {Helton}, {Ji},
  {Kumari}, {Lyu}, {Nelson}, {Parlanti}, {Perna}, {Sandles}, {Scholtz},
  {Suess}, {Topping}, {{\"U}bler}, {Wallace}, \& {Whitler}}]{Bunker2023}
{Bunker}, A.~J., {Saxena}, A., {Cameron}, A.~J., {et~al.} 2023, \aap, 677, A88

\bibitem[{{Cameron} {et~al.}(2023){Cameron}, {Saxena}, {Bunker}, {D'Eugenio},
  {Carniani}, {Maiolino}, {Curtis-Lake}, {Ferruit}, {Jakobsen}, {Arribas},
  {Bonaventura}, {Charlot}, {Chevallard}, {Curti}, {Looser}, {Maseda}, {Rawle},
  {Rodr{\'\i}guez Del Pino}, {Smit}, {{\"U}bler}, {Willott}, {Witstok},
  {Egami}, {Eisenstein}, {Johnson}, {Hainline}, {Rieke}, {Robertson}, {Stark},
  {Tacchella}, {Williams}, {Willmer}, {Bhatawdekar}, {Bowler}, {Boyett},
  {Circosta}, {Helton}, {Jones}, {Kumari}, {Ji}, {Nelson}, {Parlanti},
  {Sandles}, {Scholtz}, \& {Sun}}]{Cameron2023}
{Cameron}, A.~J., {Saxena}, A., {Bunker}, A.~J., {et~al.} 2023, \aap, 677, A115

\bibitem[{{Carniani} {et~al.}(2024){Carniani}, {Hainline}, {D'Eugenio},
  {Eisenstein}, {Jakobsen}, {Witstok}, {Johnson}, {Chevallard}, {Maiolino},
  {Helton}, {Willott}, {Robertson}, {Alberts}, {Arribas}, {Baker},
  {Bhatawdekar}, {Boyett}, {Bunker}, {Cameron}, {Cargile}, {Charlot}, {Curti},
  {Curtis-Lake}, {Egami}, {Giardino}, {Isaak}, {Ji}, {Jones}, {Maseda},
  {Parlanti}, {Rawle}, {Rieke}, {Rieke}, {Rodr{\'\i}guez Del Pino}, {Saxena},
  {Scholtz}, {Smit}, {Sun}, {Tacchella}, {{\"U}bler}, {Venturi}, {Williams}, \&
  {Willmer}}]{Carniani2024}
{Carniani}, S., {Hainline}, K., {D'Eugenio}, F., {et~al.} 2024, arXiv e-prints,
  arXiv:2405.18485

\bibitem[{{Castellano} {et~al.}(2023){Castellano}, {Fontana}, {Treu}, {Merlin},
  {Santini}, {Bergamini}, {Grillo}, {Rosati}, {Acebron}, {Leethochawalit},
  {Paris}, {Bonchi}, {Belfiori}, {Calabr{\`o}}, {Correnti}, {Nonino},
  {Polenta}, {Trenti}, {Boyett}, {Brammer}, {Broadhurst}, {Caminha}, {Chen},
  {Filippenko}, {Fortuni}, {Glazebrook}, {Mascia}, {Mason}, {Menci},
  {Meneghetti}, {Mercurio}, {Metha}, {Morishita}, {Nanayakkara}, {Pentericci},
  {Roberts-Borsani}, {Roy}, {Vanzella}, {Vulcani}, {Yang}, \&
  {Wang}}]{Castellano2023}
{Castellano}, M., {Fontana}, A., {Treu}, T., {et~al.} 2023, \apjl, 948, L14

\bibitem[{{Castellano} {et~al.}(2022){Castellano}, {Fontana}, {Treu},
  {Santini}, {Merlin}, {Leethochawalit}, {Trenti}, {Vanzella}, {Mestric},
  {Bonchi}, {Belfiori}, {Nonino}, {Paris}, {Polenta}, {Roberts-Borsani},
  {Boyett}, {Brada{\v{c}}}, {Calabr{\`o}}, {Glazebrook}, {Grillo}, {Mascia},
  {Mason}, {Mercurio}, {Morishita}, {Nanayakkara}, {Pentericci}, {Rosati},
  {Vulcani}, {Wang}, \& {Yang}}]{Castellano2022}
{Castellano}, M., {Fontana}, A., {Treu}, T., {et~al.} 2022, \apjl, 938, L15

\bibitem[{{Castellano} {et~al.}(2024){Castellano}, {Napolitano}, {Fontana},
  {Roberts-Borsani}, {Treu}, {Vanzella}, {Zavala}, {Arrabal Haro},
  {Calabr{\`o}}, {Llerena}, {Mascia}, {Merlin}, {Paris}, {Pentericci},
  {Santini}, {Bakx}, {Bergamini}, {Cupani}, {Dickinson}, {Filippenko},
  {Glazebrook}, {Grillo}, {Kelly}, {Malkan}, {Mason}, {Morishita},
  {Nanayakkara}, {Rosati}, {Sani}, {Wang}, \& {Yoon}}]{Castellano2024}
{Castellano}, M., {Napolitano}, L., {Fontana}, A., {et~al.} 2024, arXiv
  e-prints, arXiv:2403.10238

\bibitem[{{Chabrier}(2003)}]{Chabrier2003}
{Chabrier}, G. 2003, \pasp, 115, 763

\bibitem[{{Chen} {et~al.}(2022){Chen}, {Kelly}, {Treu}, {Wang},
  {Roberts-Borsani}, {Keen}, {Windhorst}, {Zhou}, {Bradac}, {Brammer},
  {Strait}, {Broadhurst}, {Diego}, {Frye}, {Meena}, {Zitrin}, {Pascale},
  {Castellano}, {Marchesini}, {Morishita}, \& {Yang}}]{Chen2022}
{Chen}, W., {Kelly}, P.~L., {Treu}, T., {et~al.} 2022, \apjl, 940, L54

\bibitem[{{Chisholm} {et~al.}(2024){Chisholm}, {Berg}, {Endsley}, {Gazagnes},
  {Richardson}, {Lambrides}, {Greene}, {Finkelstein}, {Flury}, {Guseva},
  {Henry}, {Hutchison}, {Izotov}, {Marques-Chaves}, {Oesch}, {Papovich},
  {Saldana-Lopez}, {Schaerer}, \& {Stephenson}}]{Chisholm2024}
{Chisholm}, J., {Berg}, D.~A., {Endsley}, R., {et~al.} 2024, arXiv e-prints,
  arXiv:2402.18643

\bibitem[{{Curti} {et~al.}(2023){Curti}, {D'Eugenio}, {Carniani}, {Maiolino},
  {Sandles}, {Witstok}, {Baker}, {Bennett}, {Piotrowska}, {Tacchella},
  {Charlot}, {Nakajima}, {Maheson}, {Mannucci}, {Amiri}, {Arribas}, {Belfiore},
  {Bonaventura}, {Bunker}, {Chevallard}, {Cresci}, {Curtis-Lake},
  {Hayden-Pawson}, {Jones}, {Kumari}, {Laseter}, {Looser}, {Marconi}, {Maseda},
  {Scholtz}, {Smit}, {{\"U}bler}, \& {Wallace}}]{Curti2023}
{Curti}, M., {D'Eugenio}, F., {Carniani}, S., {et~al.} 2023, \mnras, 518, 425

\bibitem[{{Curtis-Lake} {et~al.}(2023){Curtis-Lake}, {Carniani}, {Cameron},
  {Charlot}, {Jakobsen}, {Maiolino}, {Bunker}, {Witstok}, {Smit}, {Chevallard},
  {Willott}, {Ferruit}, {Arribas}, {Bonaventura}, {Curti}, {D'Eugenio},
  {Franx}, {Giardino}, {Looser}, {L{\"u}tzgendorf}, {Maseda}, {Rawle}, {Rix},
  {Rodr{\'\i}guez del Pino}, {{\"U}bler}, {Sirianni}, {Dressler}, {Egami},
  {Eisenstein}, {Endsley}, {Hainline}, {Hausen}, {Johnson}, {Rieke},
  {Robertson}, {Shivaei}, {Stark}, {Tacchella}, {Williams}, {Willmer},
  {Bhatawdekar}, {Bowler}, {Boyett}, {Chen}, {de Graaff}, {Helton}, {Hviding},
  {Jones}, {Kumari}, {Lyu}, {Nelson}, {Perna}, {Sandles}, {Saxena}, {Suess},
  {Sun}, {Topping}, {Wallace}, \& {Whitler}}]{Curtis-Lake2023}
{Curtis-Lake}, E., {Carniani}, S., {Cameron}, A., {et~al.} 2023, Nature
  Astronomy, 7, 622

\bibitem[{{Dressler} {et~al.}(2023){Dressler}, {Vulcani}, {Treu}, {Rieke},
  {Burns}, {Calabr{\`o}}, {Bonchi}, {Castellano}, {Fontana}, {Leethochawalit},
  {Mason}, {Merlin}, {Morishita}, {Paris}, {Bradac}, {Mercurio}, {Nanayakkara},
  {Poggianti}, {Santini}, {Wang}, {Misselt}, {Stark}, \&
  {Willmer}}]{Dressler2023}
{Dressler}, A., {Vulcani}, B., {Treu}, T., {et~al.} 2023, \apjl, 947, L27

\bibitem[{{Fern{\'a}ndez} {et~al.}(2024){Fern{\'a}ndez}, {Amor{\'\i}n},
  {Firpo}, \& {Morisset}}]{Fernandez2024}
{Fern{\'a}ndez}, V., {Amor{\'\i}n}, R., {Firpo}, V., \& {Morisset}, C. 2024,
  arXiv e-prints, arXiv:2405.15072

\bibitem[{{Ferruit} {et~al.}(2022){Ferruit}, {Jakobsen}, {Giardino}, {Rawle},
  {Alves de Oliveira}, {Arribas}, {Beck}, {Birkmann}, {B{\"o}ker}, {Bunker},
  {Charlot}, {de Marchi}, {Franx}, {Henry}, {Karakla}, {Kassin}, {Kumari},
  {L{\'o}pez-Caniego}, {L{\"u}tzgendorf}, {Maiolino}, {Manjavacas}, {Marston},
  {Moseley}, {Muzerolle}, {Pirzkal}, {Rauscher}, {Rix}, {Sabbi}, {Sirianni},
  {te Plate}, {Valenti}, {Willott}, \& {Zeidler}}]{Ferruit2022}
{Ferruit}, P., {Jakobsen}, P., {Giardino}, G., {et~al.} 2022, \aap, 661, A81

\bibitem[{{Gardner} {et~al.}(2023){Gardner}, {Mather}, {Abbott}, {Abell},
  {Abernathy}, {Abney}, {Abraham}, {Abraham}, {Abul-Huda}, {Acton}, {Adams},
  {Adams}, {Adler}, {Adriaensen}, {Aguilar}, {Ahmed}, {Ahmed}, {Ahmed},
  {Albat}, {Albert}, {Alberts}, {Aldridge}, {Allen}, {Allen}, {Altenburg},
  {Altunc}, {Alvarez}, {{\'A}lvarez-M{\'a}rquez}, {Alves de Oliveira},
  {Ambrose}, {Anandakrishnan}, {Andersen}, {Anderson}, {Anderson}, {Anderson},
  {Anderson}, {Aprea}, {Archer}, {Arenberg}, {Argyriou}, {Arribas}, {Artigau},
  {Arvai}, {Atcheson}, {Atkinson}, {Averbukh}, {Aymergen}, {Bacinski},
  {Baggett}, {Bagnasco}, {Baker}, {Balzano}, {Banks}, {Baran}, {Barker},
  {Barrett}, {Barringer}, {Barto}, {Bast}, {Baudoz}, {Baum}, {Beatty},
  {Beaulieu}, {Bechtold}, {Beck}, {Beddard}, {Beichman}, {Bellagama}, {Bely},
  {Berger}, {Bergeron}, {Bernier}, {Bertch}, {Beskow}, {Betz}, {Biagetti},
  {Birkmann}, {Bjorklund}, {Blackwood}, {Blazek}, {Blossfeld}, {Bluth},
  {Boccaletti}, {Boegner}, {Bohlin}, {Boia}, {B{\"o}ker}, {Bonaventura},
  {Bond}, {Bosley}, {Boucarut}, {Bouchet}, {Bouwman}, {Bower}, {Bowers},
  {Bowers}, {Boyce}, {Boyer}, {Boyer}, {Boyer}, {Boyer}, {Bradley}, {Brady},
  {Brandl}, {Brannen}, {Breda}, {Bremmer}, {Brennan}, {Bresnahan}, {Bright},
  {Broiles}, {Bromenschenkel}, {Brooks}, {Brooks}, {Brown}, {Brown}, {Brown},
  {Bruce}, {Bryson}, {Bujanda}, {Bullock}, {Bunker}, {Bureo}, {Burt}, {Bush},
  {Bushouse}, {Bussman}, {Cabaud}, {Cale}, {Calhoon}, {Calvani}, {Canipe},
  {Caputo}, {Cara}, {Carey}, {Case}, {Cesari}, {Cetorelli}, {Chance},
  {Chandler}, {Chaney}, {Chapman}, {Charlot}, {Chayer}, {Cheezum}, {Chen},
  {Chen}, {Cherinka}, {Chichester}, {Chilton}, {Chittiraibalan}, {Clampin},
  {Clark}, {Clark}, {Clark}, {Claybrooks}, {Cleveland}, {Cohen}, {Cohen},
  {Col{\'o}n}, {Coleman}, {Colina}, {Comber}, {Comeau}, {Comer}, {Conde Reis},
  {Connolly}, {Conroy}, {Contos}, {Contreras}, {Cook}, {Cooper}, {Cooper},
  {Correia}, {Correnti}, {Cossou}, {Costanza}, {Coulais}, {Cox}, {Coyle},
  {Cracraft}, {Crew}, {Curtis}, {Cusveller}, {Da Costa Maciel}, {Dailey},
  {Daugeron}, {Davidson}, {Davies}, {Davis}, {Davis}, {Day}, {de Chambure}, {de
  Jong}, {De Marchi}, {Dean}, {Decker}, {Delisa}, {Dell}, {Dellagatta},
  {Dembinska}, {Demosthenes}, {Dencheva}, {Deneu}, {DePriest}, {Deschenes},
  {Dethienne}, {Detre}, {Diaz}, {Dicken}, {DiFelice}, {Dillman}, {Disharoon},
  {Dixon}, {Doggett}, {Dominguez}, {Donaldson}, {Doria-Warner}, {Santos},
  {Doty}, {Douglas}, {Doyon}, {Dressler}, {Driggers}, {Driggers}, {Dunn},
  {DuPrie}, {Dupuis}, {Durning}, {Dutta}, {Earl}, {Eccleston}, {Ecobichon},
  {Egami}, {Ehrenwinkler}, {Eisenhamer}, {Eisenhower}, {Eisenstein}, {El
  Hamel}, {Elie}, {Elliott}, {Elliott}, {Engesser}, {Espinoza}, {Etienne},
  {Etxaluze}, {Evans}, {Fabreguettes}, {Falcolini}, {Falini}, {Fatig},
  {Feeney}, {Feinberg}, {Fels}, {Ferdous}, {Ferguson}, {Ferrarese}, {Ferreira},
  {Ferruit}, {Ferry}, {Filippazzo}, {Firre}, {Fix}, {Flagey}, {Flanagan},
  {Fleming}, {Florian}, {Flynn}, {Foiadelli}, {Fontaine}, {Fontanella},
  {Forshay}, {Fortner}, {Fox}, {Framarini}, {Francisco}, {Franck}, {Franx},
  {Franz}, {Friedman}, {Friend}, {Frost}, {Fu}, {Fullerton}, {Gaillard},
  {Galkin}, {Gallagher}, {Galyer}, {Garc{\'\i}a Mar{\'\i}n}, {Gardner},
  {Garland}, {Garrett}, {Gasman}, {G{\'a}sp{\'a}r}, {Gastaud}, {Gaudreau},
  {Gauthier}, {Geers}, {Geithner}, {Gennaro}, {Gerber}, {Gereau}, {Giampaoli},
  {Giardino}, {Gibbons}, {Gilbert}, {Gilman}, {Girard}, {Giuliano}, {Gkountis},
  {Glasse}, {Glassmire}, {Glauser}, {Glazer}, {Goldberg}, {Golimowski},
  {Gonzaga}, {Gordon}, {Gordon}, {Goudfrooij}, {Gough}, {Graham}, {Grau},
  {Green}, {Greene}, {Greene}, {Greenfield}, {Greenhouse}, {Greve}, {Greville},
  {Grimaldi}, {Groe}, {Groebner}, {Grumm}, {Grundy}, {G{\"u}del}, {Guillard},
  {Guldalian}, {Gunn}, {Gurule}, {Gutman}, {Guy}, {Guyot}, {Hack}, {Haderlein},
  {Hagan}, {Hagedorn}, {Hainline}, {Haley}, {Hami}, {Hamilton}, {Hammann},
  {Hammel}, {Hanley}, {Hansen}, {Hardy}, {Harnisch}, {Harr}, {Harris}, {Hart},
  {Hartig}, {Hasan}, {Hashim}, {Hashimoto}, {Haskins}, {Hawkins}, {Hayden},
  {Hayden}, {Healy}, {Hecht}, {Heeg}, {Hejal}, {Helm}, {Hengemihle}, {Henning},
  {Henry}, {Henry}, {Henshaw}, {Hernandez}, {Herrington}, {Heske}, {Hesman},
  {Hickey}, {Hilbert}, {Hines}, {Hinz}, {Hirsch}, {Hitcho}, {Hodapp}, {Hodge},
  {Hoffman}, {Holfeltz}, {Holler}, {Hoppa}, {Horner}, {Howard}, {Howard},
  {Huber}, {Hunkeler}, {Hunter}, {Hunter}, {Hurd}, {Hurst}, {Hutchings},
  {Hylan}, {Ignat}, {Illingworth}, {Irish}, {Isaacs}, {Jackson}, {Jaffe},
  {Jahic}, {Jahromi}, {Jakobsen}, {James}, {James}, {James}, {Jamieson},
  {Jandra}, {Jayawardhana}, {Jedrzejewski}, {Jeffers}, {Jensen}, {Joanne},
  {Johns}, {Johnson}, {Johnson}, {Johnson}, {Johnson}, {Johnson}, {Johnson},
  {Johnstone}, {Jollet}, {Jones}, {Jones}, {Jones}, {Jones}, {Jones}, {Jordan},
  {Jordan}, {Jue}, {Jurkowski}, {Justis}, {Justtanont}, {Kaleida}, {Kalirai},
  {Kalmanson}, {Kaltenegger}, {Kammerer}, {Kan}, {Kanarek}, {Kao}, {Karakla},
  {Karl}, {Kassin}, {Kauffman}, {Kavanagh}, {Kelley}, {Kelly}, {Kendrew},
  {Kennedy}, {Kenny}, {Keski-Kuha}, {Keyes}, {Khan}, {Kidwell}, {Kimble},
  {King}, {King}, {Kinzel}, {Kirk}, {Kirkpatrick}, {Klaassen}, {Klingemann},
  {Klintworth}, {Knapp}, {Knight}, {Knollenberg}, {Knutsen}, {Koehler},
  {Koekemoer}, {Kofler}, {Kontson}, {Kovacs}, {Kozhurina-Platais}, {Krause},
  {Kriss}, {Krist}, {Kristoffersen}, {Krogel}, {Krueger}, {Kulp}, {Kumari},
  {Kwan}, {Kyprianou}, {Labador}, {Labiano}, {Lafreni{\`e}re}, {Lagage},
  {Laidler}, {Laine}, {Laird}, {Lajoie}, {Lallo}, {Lam}, {LaMassa}, {Lambros},
  {Lampenfield}, {Lander}, {Langston}, {Larson}, {Larson}, {LaVerghetta},
  {Law}, {Lawrence}, {Lee}, {Lee}, {Lee}, {Leisenring}, {Leveille}, {Levenson},
  {Levi}, {Levine}, {Lewis}, {Lewis}, {Lewis}, {Libralato}, {Lidon},
  {Liebrecht}, {Lightsey}, {Lilly}, {Lim}, {Lim}, {Ling}, {Link}, {Link},
  {Lipinski}, {Liu}, {Lo}, {Lobmeyer}, {Logue}, {Long}, {Long}, {Long}, {Long},
  {L{\'o}pez-Caniego}, {Lotz}, {Love-Pruitt}, {Lubskiy}, {Luers}, {Luetgens},
  {Luevano}, {Lui}, {Lund}, {Lundquist}, {Lunine}, {L{\"u}tzgendorf}, {Lynch},
  {MacDonald}, {MacDonald}, {Macias}, {Macklis}, {Maghami}, {Maharaja},
  {Maiolino}, {Makrygiannis}, {Malla}, {Malumuth}, {Manjavacas}, {Marini},
  {Marrione}, {Marston}, {Martel}, {Martin}, {Martin}, {Martinez}, {Maschmann},
  {Masci}, {Masetti}, {Maszkiewicz}, {Matthews}, {Matuskey}, {McBrayer},
  {McCarthy}, {McCaughrean}, {McClare}, {McClare}, {McCloskey}, {McClurg},
  {McCoy}, {McElwain}, {McGregor}, {McGuffey}, {McKay}, {McKenzie}, {McLean},
  {McMaster}, {McNeil}, {De Meester}, {Mehalick}, {Meixner}, {Mel{\'e}ndez},
  {Menzel}, {Menzel}, {Merz}, {Mesterharm}, {Meyer}, {Meyett}, {Meza},
  {Midwinter}, {Milam}, {Miller}, {Miller}, {Miskey}, {Misselt}, {Mitchell},
  {Mohan}, {Montoya}, {Moran}, {Morishita}, {Moro-Mart{\'\i}n}, {Morrison},
  {Morrison}, {Morse}, {Moschos}, {Moseley}, {Mosier}, {Mosner}, {Mountain},
  {Muckenthaler}, {Mueller}, {Mueller}, {Muhiem}, {M{\"u}hlmann}, {Mullally},
  {Mullen}, {Munger}, {Murphy}, {Murray}, {Muzerolle}, {Mycroft}, {Myers},
  {Myers}, {Myers}, {Myers}, {Myrick}, {Nagle}, {Nayak}, {Naylor}, {Neff},
  {Nelan}, {Nella}, {Nguyen}, {Nguyen}, {Nickson}, {Nidhiry}, {Niedner},
  {Nieto-Santisteban}, {Nikolov}, {Nishisaka}, {Noriega-Crespo}, {Nota},
  {O'Mara}, {Oboryshko}, {O'Brien}, {Ochs}, {Offenberg}, {Ogle}, {Ohl},
  {Olmsted}, {Osborne}, {O'Shaughnessy}, {{\"O}stlin}, {O'Sullivan}, {Otor},
  {Ottens}, {Ouellette}, {Outlaw}, {Owens}, {Pacifici}, {Page}, {Paranilam},
  {Park}, {Parrish}, {Paschal}, {Patapis}, {Patel}, {Patrick}, {Pattishall},
  {Paul}, {Paul}, {Pauly}, {Pavlovsky}, {Pe{\~n}a-Guerrero}, {Pedder}, {Peek},
  {Pelham}, {Penanen}, {Perriello}, {Perrin}, {Perrine}, {Perrygo}, {Peslier},
  {Petach}, {Peterson}, {Pfarr}, {Pierson}, {Pietraszkiewicz}, {Pilchen},
  {Pipher}, {Pirzkal}, {Pitman}, {Player}, {Plesha}, {Plitzke}, {Pohner},
  {Poletis}, {Pollizzi}, {Polster}, {Pontius}, {Pontoppidan}, {Porges},
  {Potter}, {Prescott}, {Proffitt}, {Pueyo}, {Quispe Neira}, {Radich}, {Rager},
  {Rameau}, {Ramey}, {Ramos Alarcon}, {Rampini}, {Rapp}, {Rashford},
  {Rauscher}, {Ravindranath}, {Rawle}, {Rawlings}, {Ray}, {Regan}, {Rehm},
  {Rehm}, {Reid}, {Reis}, {Renk}, {Reoch}, {Ressler}, {Rest}, {Reynolds},
  {Richon}, {Richon}, {Ridgaway}, {Riedel}, {Rieke}, {Rieke}, {Rifelli},
  {Rigby}, {Riggs}, {Ringel}, {Ritchie}, {Rix}, {Robberto}, {Robinson},
  {Robinson}, {Robinson}, {Rock}, {Rodriguez}, {Rodr{\'\i}guez del Pino},
  {Roellig}, {Rohrbach}, {Roman}, {Romelfanger}, {Romo}, {Rosales}, {Rose},
  {Roteliuk}, {Roth}, {Rothwell}, {Rouzaud}, {Rowe}, {Rowlands}, {Roy},
  {Royer}, {Rui}, {Rumler}, {Rumpl}, {Russ}, {Ryan}, {Ryan}, {Saad}, {Sabata},
  {Sabatino}, {Sabbi}, {Sabelhaus}, {Sabia}, {Sahu}, {Saif}, {Salvignol},
  {Samara-Ratna}, {Samuelson}, {Sanders}, {Sappington}, {Sargent}, {Sauer},
  {Savadkin}, {Sawicki}, {Schappell}, {Scheffer}, {Scheithauer}, {Scherer},
  {Schiff}, {Schlawin}, {Schmeitzky}, {Schmitz}, {Schmude}, {Schneider},
  {Schreiber}, {Schroeven-Deceuninck}, {Schultz}, {Schwab}, {Schwartz},
  {Scoccimarro}, {Scott}, {Scott}, {Seaton}, {Seely}, {Seery}, {Seidleck},
  {Sembach}, {Shanahan}, {Shaughnessy}, {Shaw}, {Shay}, {Sheehan}, {Sheth},
  {Shih}, {Shivaei}, {Siegel}, {Sienkiewicz}, {Simmons}, {Simon}, {Sirianni},
  {Sivaramakrishnan}, {Slade}, {Sloan}, {Slocum}, {Slowinski}, {Smith},
  {Smith}, {Smith}, {Smith}, {Smith}, {Smith}, {Smolik}, {Soderblom}, {Sohn},
  {Sokol}, {Sonneborn}, {Sontag}, {Sooy}, {Soummer}, {Southwood}, {Spain},
  {Sparmo}, {Speer}, {Spencer}, {Sprofera}, {Stallcup}, {Stanley},
  {Stansberry}, {Stark}, {Starr}, {Stassi}, {Steck}, {Steeley}, {Stephens},
  {Stephenson}, {Stewart}, {Stiavelli}, {}, {Strada}, {Straughn}, {Streetman},
  {Strickland}, {Strobele}, {Stuhlinger}, {Stys}, {Such}, {Sukhatme},
  {Sullivan}, {Sullivan}, {Sumner}, {Sun}, {Sunnquist}, {Swade}, {Swam},
  {Swenton}, {Swoish}, {Tam Litten}, {Tamas}, {Tao}, {Taylor}, {Taylor}, {te
  Plate}, {Van Tea}, {Teague}, {Telfer}, {Temim}, {Texter}, {Thatte},
  {Thompson}, {Thompson}, {Thomson}, {Thronson}, {Tierney}, {Tikkanen},
  {Tinnin}, {Tippet}, {Todd}, {Tran}, {Trauger}, {Trejo}, {Vinh Truong},
  {Tsukamoto}, {Tufail}, {Tumlinson}, {Tustain}, {Tyra}, {Ubeda}, {Underwood},
  {Uzzo}, {Vaclavik}, {Valenduc}, {Valenti}, {Van Campen}, {van de Wetering},
  {Van Der Marel}, {van Haarlem}, {Vandenbussche}, {van Dishoeck},
  {Vanterpool}, {Vernoy}, {Vila Costas}, {Volk}, {Voorzaat}, {Voyton}, {Vydra},
  {Waddy}, {Waelkens}, {Wahlgren}, {Walker}, {Wander}, {Warfield}, {Warner},
  {Wasiak}, {Wasiak}, {Wehner}, {Weiler}, {Weilert}, {Weiss}, {Wells}, {Welty},
  {Wheate}, {Wheeler}, {White}, {Whitehouse}, {Whiteleather}, {Whitman},
  {Williams}, {Willmer}, {Willott}, {Willoughby}, {Wilson}, {Wilson}, {Wilson},
  {Windhorst}, {Wislowski}, {Wolfe}, {Wolfe}, {Wolff}, {Wondel}, {Woo},
  {Woods}, {Worden}, {Workman}, {Wright}, {Wu}, {Wu}, {Wun}, {Wymer},
  {Yadetie}, {Yan}, {Yang}, {Yates}, {Yeager}, {Yerger}, {Young}, {Young},
  {Yu}, {Yu}, {Zak}, {Zeidler}, {Zepp}, {Zhou}, {Zincke}, {Zonak}, \&
  {Zondag}}]{Gardner2023}
{Gardner}, J.~P., {Mather}, J.~C., {Abbott}, R., {et~al.} 2023, \pasp, 135,
  068001

\bibitem[{{Glazebrook} {et~al.}(2023){Glazebrook}, {Nanayakkara}, {Jacobs},
  {Leethochawalit}, {Calabr{\`o}}, {Bonchi}, {Castellano}, {Fontana}, {Mason},
  {Merlin}, {Morishita}, {Paris}, {Trenti}, {Treu}, {Santini}, {Wang},
  {Boyett}, {Bradac}, {Brammer}, {Jones}, {Marchesini}, {Nonino}, \&
  {Vulcani}}]{Glazebrook2023}
{Glazebrook}, K., {Nanayakkara}, T., {Jacobs}, C., {et~al.} 2023, \apjl, 947,
  L25

\bibitem[{{Harikane} {et~al.}(2024){Harikane}, {Nakajima}, {Ouchi}, {Umeda},
  {Isobe}, {Ono}, {Xu}, \& {Zhang}}]{Harikane2023}
{Harikane}, Y., {Nakajima}, K., {Ouchi}, M., {et~al.} 2024, \apj, 960, 56

\bibitem[{{Isobe} {et~al.}(2023){Isobe}, {Ouchi}, {Nakajima}, {Harikane},
  {Ono}, {Xu}, {Zhang}, \& {Umeda}}]{Isobe2023}
{Isobe}, Y., {Ouchi}, M., {Nakajima}, K., {et~al.} 2023, \apj, 956, 139

\bibitem[{{Jacobs} {et~al.}(2023){Jacobs}, {Glazebrook}, {Calabr{\`o}}, {Treu},
  {Nannayakkara}, {Jones}, {Merlin}, {Abraham}, {Stevens}, {Vulcani}, {Yang},
  {Bonchi}, {Boyett}, {Brada{\v{c}}}, {Castellano}, {Fontana}, {Marchesini},
  {Malkan}, {Mason}, {Morishita}, {Paris}, {Santini}, {Trenti}, \&
  {Wang}}]{Jacobs2023}
{Jacobs}, C., {Glazebrook}, K., {Calabr{\`o}}, A., {et~al.} 2023, \apjl, 948,
  L13

\bibitem[{{Jakobsen} {et~al.}(2022){Jakobsen}, {Ferruit}, {Alves de Oliveira},
  {Arribas}, {Bagnasco}, {Barho}, {Beck}, {Birkmann}, {B{\"o}ker}, {Bunker},
  {Charlot}, {de Jong}, {de Marchi}, {Ehrenwinkler}, {Falcolini}, {Fels},
  {Franx}, {Franz}, {Funke}, {Giardino}, {Gnata}, {Holota}, {Honnen}, {Jensen},
  {Jentsch}, {Johnson}, {Jollet}, {Karl}, {Kling}, {K{\"o}hler}, {Kolm},
  {Kumari}, {Lander}, {Lemke}, {L{\'o}pez-Caniego}, {L{\"u}tzgendorf},
  {Maiolino}, {Manjavacas}, {Marston}, {Maschmann}, {Maurer}, {Messerschmidt},
  {Moseley}, {Mosner}, {Mott}, {Muzerolle}, {Pirzkal}, {Pittet}, {Plitzke},
  {Posselt}, {Rapp}, {Rauscher}, {Rawle}, {Rix}, {R{\"o}del}, {Rumler},
  {Sabbi}, {Salvignol}, {Schmid}, {Sirianni}, {Smith}, {Strada}, {te Plate},
  {Valenti}, {Wettemann}, {Wiehe}, {Wiesmayer}, {Willott}, {Wright}, {Zeidler},
  \& {Zincke}}]{Jakobsen2022}
{Jakobsen}, P., {Ferruit}, P., {Alves de Oliveira}, C., {et~al.} 2022, \aap,
  661, A80

\bibitem[{{Jones} {et~al.}(2023){Jones}, {Sanders}, {Chen}, {Wang},
  {Morishita}, {Roberts-Borsani}, {Treu}, {Dressler}, {Merlin}, {Paris},
  {Santini}, {Bergamini}, {Henry}, {Huntzinger}, {Nanayakkara}, {Boyett},
  {Bradac}, {Brammer}, {Calabr{\'o}}, {Glazebrook}, {Grasha}, {Mascia},
  {Pentericci}, {Trenti}, \& {Vulcani}}]{Jones2023}
{Jones}, T., {Sanders}, R., {Chen}, Y., {et~al.} 2023, \apjl, 951, L17

\bibitem[{{Larson} {et~al.}(2023){Larson}, {Finkelstein}, {Kocevski},
  {Hutchison}, {Trump}, {Arrabal Haro}, {Bromm}, {Cleri}, {Dickinson},
  {Fujimoto}, {Kartaltepe}, {Koekemoer}, {Papovich}, {Pirzkal}, {Tacchella},
  {Zavala}, {Bagley}, {Behroozi}, {Champagne}, {Cole}, {Jung}, {Morales},
  {Yang}, {Zhang}, {Zitrin}, {Amor{\'\i}n}, {Burgarella}, {Casey}, {Ch{\'a}vez
  Ortiz}, {Cox}, {Chworowsky}, {Fontana}, {Gawiser}, {Grazian}, {Grogin},
  {Harish}, {Hathi}, {Hirschmann}, {Holwerda}, {Juneau}, {Leung}, {Lucas},
  {McGrath}, {P{\'e}rez-Gonz{\'a}lez}, {Rigby}, {Seill{\'e}}, {Simons}, {de La
  Vega}, {Weiner}, {Wilkins}, {Yung}, \& {Ceers Team}}]{Larson2023}
{Larson}, R.~L., {Finkelstein}, S.~L., {Kocevski}, D.~D., {et~al.} 2023, \apjl,
  953, L29

\bibitem[{{Laseter} {et~al.}(2024){Laseter}, {Maseda}, {Curti}, {Maiolino},
  {D'Eugenio}, {Cameron}, {Looser}, {Arribas}, {Baker}, {Bhatawdekar},
  {Boyett}, {Bunker}, {Carniani}, {Charlot}, {Chevallard}, {Curtis-lake},
  {Egami}, {Eisenstein}, {Hainline}, {Hausen}, {Ji}, {Kumari}, {Perna},
  {Rawle}, {Rix}, {Robertson}, {Rodr{\'\i}guez Del Pino}, {Sandles}, {Scholtz},
  {Smit}, {Tacchella}, {{\"U}bler}, {Williams}, {Willott}, \&
  {Witstok}}]{Laseter2024}
{Laseter}, I.~H., {Maseda}, M.~V., {Curti}, M., {et~al.} 2024, \aap, 681, A70

\bibitem[{{Leethochawalit} {et~al.}(2023){Leethochawalit}, {Trenti}, {Santini},
  {Yang}, {Merlin}, {Castellano}, {Fontana}, {Treu}, {Mason}, {Glazebrook},
  {Jones}, {Vulcani}, {Nanayakkara}, {Marchesini}, {Mascia}, {Morishita},
  {Roberts-Borsani}, {Bonchi}, {Paris}, {Boyett}, {Strait}, {Calabr{\`o}},
  {Pentericci}, {Bradac}, {Wang}, \& {Scarlata}}]{Leethochawalit2023}
{Leethochawalit}, N., {Trenti}, M., {Santini}, P., {et~al.} 2023, \apjl, 942,
  L26

\bibitem[{{Maiolino} {et~al.}(2024){Maiolino}, {Scholtz}, {Witstok},
  {Carniani}, {D'Eugenio}, {de Graaff}, {{\"U}bler}, {Tacchella},
  {Curtis-Lake}, {Arribas}, {Bunker}, {Charlot}, {Chevallard}, {Curti},
  {Looser}, {Maseda}, {Rawle}, {Rodr{\'\i}guez del Pino}, {Willott}, {Egami},
  {Eisenstein}, {Hainline}, {Robertson}, {Williams}, {Willmer}, {Baker},
  {Boyett}, {DeCoursey}, {Fabian}, {Helton}, {Ji}, {Jones}, {Kumari},
  {Laporte}, {Nelson}, {Perna}, {Sandles}, {Shivaei}, \& {Sun}}]{Maiolino2024}
{Maiolino}, R., {Scholtz}, J., {Witstok}, J., {et~al.} 2024, \nat, 627, 59

\bibitem[{{Marchesini} {et~al.}(2023){Marchesini}, {Brammer}, {Morishita},
  {Bergamini}, {Wang}, {Bradac}, {Roberts-Borsani}, {Strait}, {Treu},
  {Fontana}, {Jones}, {Santini}, {Vulcani}, {Acebron}, {Calabr{\`o}},
  {Castellano}, {Glazebrook}, {Grillo}, {Mercurio}, {Nanayakkara}, {Rosati},
  {Tubthong}, \& {Vanzella}}]{Marchesini2023}
{Marchesini}, D., {Brammer}, G., {Morishita}, T., {et~al.} 2023, \apjl, 942,
  L25

\bibitem[{Markwardt(2009)}]{markwardt2009}
Markwardt, C.~B. 2009, Non-linear Least Squares Fitting in IDL with MPFIT

\bibitem[{{Mascia} {et~al.}(2023){Mascia}, {Pentericci}, {Calabr{\`o}}, {Treu},
  {Santini}, {Yang}, {Napolitano}, {Roberts-Borsani}, {Bergamini}, {Grillo},
  {Rosati}, {Vulcani}, {Castellano}, {Boyett}, {Fontana}, {Glazebrook},
  {Henry}, {Mason}, {Merlin}, {Morishita}, {Nanayakkara}, {Paris}, {Roy},
  {Williams}, {Wang}, {Brammer}, {Brada{\v{c}}}, {Chen}, {Kelly}, {Koekemoer},
  {Trenti}, \& {Windhorst}}]{Mascia2023}
{Mascia}, S., {Pentericci}, L., {Calabr{\`o}}, A., {et~al.} 2023, \aap, 672,
  A155

\bibitem[{{Morishita} {et~al.}(2023){Morishita}, {Roberts-Borsani}, {Treu},
  {Brammer}, {Mason}, {Trenti}, {Vulcani}, {Wang}, {Acebron}, {Bah{\'e}},
  {Bergamini}, {Boyett}, {Bradac}, {Calabr{\`o}}, {Castellano}, {Chen}, {De
  Lucia}, {Filippenko}, {Fontana}, {Glazebrook}, {Grillo}, {Henry}, {Jones},
  {Kelly}, {Koekemoer}, {Leethochawalit}, {Lu}, {Marchesini}, {Mascia},
  {Mercurio}, {Merlin}, {Metha}, {Nanayakkara}, {Nonino}, {Paris},
  {Pentericci}, {Rosati}, {Santini}, {Strait}, {Vanzella}, {Windhorst}, \&
  {Xie}}]{Morishita2023}
{Morishita}, T., {Roberts-Borsani}, G., {Treu}, T., {et~al.} 2023, \apjl, 947,
  L24

\bibitem[{{Nakajima} {et~al.}(2023){Nakajima}, {Ouchi}, {Isobe}, {Harikane},
  {Zhang}, {Ono}, {Umeda}, \& {Oguri}}]{Nakajima2023}
{Nakajima}, K., {Ouchi}, M., {Isobe}, Y., {et~al.} 2023, arXiv e-prints,
  arXiv:2301.12825

\bibitem[{{Nanayakkara} {et~al.}(2023){Nanayakkara}, {Glazebrook}, {Jacobs},
  {Bonchi}, {Castellano}, {Fontana}, {Mason}, {Merlin}, {Morishita}, {Paris},
  {Trenti}, {Treu}, {Calabr{\`o}}, {Boyett}, {Bradac}, {Leethochawalit},
  {Marchesini}, {Santini}, {Strait}, {Vanzella}, {Vulcani}, {Wang}, \&
  {Yang}}]{Nanayakkara2023}
{Nanayakkara}, T., {Glazebrook}, K., {Jacobs}, C., {et~al.} 2023, \apjl, 947,
  L26

\bibitem[{{Napolitano} {et~al.}(2024){Napolitano}, {Pentericci}, {Santini},
  {Calabr{\`o}}, {Mascia}, {Llerena}, {Castellano}, {Dickinson}, {Finkelstein},
  {Amorin}, {Arrabal Haro}, {Bagley}, {Bhatawdekar}, {Cleri}, {Davis},
  {Gardner}, {Gawiser}, {Giavalisco}, {Hathi}, {Hu}, {Jung}, {Kartaltepe},
  {Koekemoer}, {Merlin}, {Mobasher}, {Papovich}, {Park}, {Pirzkal}, {Trump},
  {Wilkins}, \& {Yung}}]{Napolitano2024}
{Napolitano}, L., {Pentericci}, L., {Santini}, P., {et~al.} 2024, arXiv
  e-prints, arXiv:2402.11220

\bibitem[{{Nonino} {et~al.}(2023){Nonino}, {Glazebrook}, {Burgasser},
  {Polenta}, {Morishita}, {Lepinzan}, {Castellano}, {Fontana}, {Merlin},
  {Bonchi}, {Paris}, {Treu}, {Vulcani}, {Wang}, {Santini}, {Vanzella},
  {Nanayakkara}, {Mercurio}, {Rosati}, {Grillo}, \& {Bradac}}]{Nonino2023}
{Nonino}, M., {Glazebrook}, K., {Burgasser}, A.~J., {et~al.} 2023, \apjl, 942,
  L29

\bibitem[{{Oke} \& {Gunn}(1983)}]{Oke1983}
{Oke}, J.~B. \& {Gunn}, J.~E. 1983, \apj, 266, 713

\bibitem[{{Planck Collaboration} {et~al.}(2020){Planck Collaboration},
  {Aghanim}, {Akrami}, {Ashdown}, {Aumont}, {Baccigalupi}, {Ballardini},
  {Banday}, {Barreiro}, {Bartolo}, {Basak}, {Battye}, {Benabed}, {Bernard},
  {Bersanelli}, {Bielewicz}, {Bock}, {Bond}, {Borrill}, {Bouchet}, {Boulanger},
  {Bucher}, {Burigana}, {Butler}, {Calabrese}, {Cardoso}, {Carron},
  {Challinor}, {Chiang}, {Chluba}, {Colombo}, {Combet}, {Contreras}, {Crill},
  {Cuttaia}, {de Bernardis}, {de Zotti}, {Delabrouille}, {Delouis}, {Di
  Valentino}, {Diego}, {Dor{\'e}}, {Douspis}, {Ducout}, {Dupac}, {Dusini},
  {Efstathiou}, {Elsner}, {En{\ss}lin}, {Eriksen}, {Fantaye}, {Farhang},
  {Fergusson}, {Fernandez-Cobos}, {Finelli}, {Forastieri}, {Frailis},
  {Fraisse}, {Franceschi}, {Frolov}, {Galeotta}, {Galli}, {Ganga},
  {G{\'e}nova-Santos}, {Gerbino}, {Ghosh}, {Gonz{\'a}lez-Nuevo}, {G{\'o}rski},
  {Gratton}, {Gruppuso}, {Gudmundsson}, {Hamann}, {Handley}, {Hansen},
  {Herranz}, {Hildebrandt}, {Hivon}, {Huang}, {Jaffe}, {Jones}, {Karakci},
  {Keih{\"a}nen}, {Keskitalo}, {Kiiveri}, {Kim}, {Kisner}, {Knox},
  {Krachmalnicoff}, {Kunz}, {Kurki-Suonio}, {Lagache}, {Lamarre}, {Lasenby},
  {Lattanzi}, {Lawrence}, {Le Jeune}, {Lemos}, {Lesgourgues}, {Levrier},
  {Lewis}, {Liguori}, {Lilje}, {Lilley}, {Lindholm}, {L{\'o}pez-Caniego},
  {Lubin}, {Ma}, {Mac{\'\i}as-P{\'e}rez}, {Maggio}, {Maino}, {Mandolesi},
  {Mangilli}, {Marcos-Caballero}, {Maris}, {Martin}, {Martinelli},
  {Mart{\'\i}nez-Gonz{\'a}lez}, {Matarrese}, {Mauri}, {McEwen}, {Meinhold},
  {Melchiorri}, {Mennella}, {Migliaccio}, {Millea}, {Mitra},
  {Miville-Desch{\^e}nes}, {Molinari}, {Montier}, {Morgante}, {Moss}, {Natoli},
  {N{\o}rgaard-Nielsen}, {Pagano}, {Paoletti}, {Partridge}, {Patanchon},
  {Peiris}, {Perrotta}, {Pettorino}, {Piacentini}, {Polastri}, {Polenta},
  {Puget}, {Rachen}, {Reinecke}, {Remazeilles}, {Renzi}, {Rocha}, {Rosset},
  {Roudier}, {Rubi{\~n}o-Mart{\'\i}n}, {Ruiz-Granados}, {Salvati}, {Sandri},
  {Savelainen}, {Scott}, {Shellard}, {Sirignano}, {Sirri}, {Spencer},
  {Sunyaev}, {Suur-Uski}, {Tauber}, {Tavagnacco}, {Tenti}, {Toffolatti},
  {Tomasi}, {Trombetti}, {Valenziano}, {Valiviita}, {Van Tent}, {Vibert},
  {Vielva}, {Villa}, {Vittorio}, {Wandelt}, {Wehus}, {White}, {White},
  {Zacchei}, \& {Zonca}}]{Planck2020}
{Planck Collaboration}, {Aghanim}, N., {Akrami}, Y., {et~al.} 2020, \aap, 641,
  A6

\bibitem[{{Prieto-Lyon} {et~al.}(2023){Prieto-Lyon}, {Mason}, {Mascia},
  {Merlin}, {Roy}, {Henry}, {Roberts-Borsani}, {Morishita}, {Wang}, {Boyett},
  {Bolan}, {Bradac}, {Castellano}, {Mercurio}, {Nanayakkara}, {Paris},
  {Pentericci}, {Scarlata}, {Trenti}, {Treu}, \& {Vanzella}}]{Prieto-Lyon2023}
{Prieto-Lyon}, G., {Mason}, C., {Mascia}, S., {et~al.} 2023, \apj, 956, 136

\bibitem[{{Reddy} {et~al.}(2023){Reddy}, {Topping}, {Sanders}, {Shapley}, \&
  {Brammer}}]{Reddy2023}
{Reddy}, N.~A., {Topping}, M.~W., {Sanders}, R.~L., {Shapley}, A.~E., \&
  {Brammer}, G. 2023, \apj, 952, 167

\bibitem[{{Richard} {et~al.}(2021){Richard}, {Claeyssens}, {Lagattuta},
  {Guaita}, {Bauer}, {Pello}, {Carton}, {Bacon}, {Soucail}, {Lyon}, {Kneib},
  {Mahler}, {Cl{\'e}ment}, {Mercier}, {Variu}, {Tamone}, {Ebeling}, {Schmidt},
  {Nanayakkara}, {Maseda}, {Weilbacher}, {Bouch{\'e}}, {Bouwens}, {Wisotzki},
  {de la Vieuville}, {Martinez}, \& {Patr{\'\i}cio}}]{Richard2021}
{Richard}, J., {Claeyssens}, A., {Lagattuta}, D., {et~al.} 2021, \aap, 646, A83

\bibitem[{{Roberts-Borsani} {et~al.}(2022){Roberts-Borsani}, {Morishita},
  {Treu}, {Brammer}, {Strait}, {Wang}, {Bradac}, {Acebron}, {Bergamini},
  {Boyett}, {Calabr{\'o}}, {Castellano}, {Fontana}, {Glazebrook}, {Grillo},
  {Henry}, {Jones}, {Malkan}, {Marchesini}, {Mascia}, {Mason}, {Mercurio},
  {Merlin}, {Nanayakkara}, {Pentericci}, {Rosati}, {Santini}, {Scarlata},
  {Trenti}, {Vanzella}, {Vulcani}, \& {Willott}}]{Roberts-Borsani2022}
{Roberts-Borsani}, G., {Morishita}, T., {Treu}, T., {et~al.} 2022, \apjl, 938,
  L13

\bibitem[{{Roberts-Borsani} {et~al.}(2023){Roberts-Borsani}, {Treu}, {Chen},
  {Morishita}, {Vanzella}, {Zitrin}, {Bergamini}, {Castellano}, {Fontana},
  {Glazebrook}, {Grillo}, {Kelly}, {Merlin}, {Nanayakkara}, {Paris}, {Rosati},
  {Yang}, {Acebron}, {Bonchi}, {Boyett}, {Brada{\v{c}}}, {Brammer},
  {Broadhurst}, {Calabr{\'o}}, {Diego}, {Dressler}, {Furtak}, {Filippenko},
  {Henry}, {Koekemoer}, {Leethochawalit}, {Malkan}, {Mason}, {Mercurio},
  {Metha}, {Pentericci}, {Pierel}, {Rieck}, {Roy}, {Santini}, {Strait},
  {Strausbaugh}, {Trenti}, {Vulcani}, {Wang}, {Wang}, \&
  {Windhorst}}]{Roberts-Borsani2023}
{Roberts-Borsani}, G., {Treu}, T., {Chen}, W., {et~al.} 2023, \nat, 618, 480

\bibitem[{{Roberts-Borsani} {et~al.}(2024){Roberts-Borsani}, {Treu}, {Shapley},
  {Fontana}, {Pentericci}, {Castellano}, {Morishita}, {Bergamini}, \&
  {Rosati}}]{Roberts-Borsani2024}
{Roberts-Borsani}, G., {Treu}, T., {Shapley}, A., {et~al.} 2024, arXiv
  e-prints, arXiv:2403.07103

\bibitem[{{Roy} {et~al.}(2023){Roy}, {Henry}, {Treu}, {Jones}, {Prieto-Lyon},
  {Mason}, {Heckman}, {Nanayakkara}, {Pentericci}, {Mascia}, {Brada{\v{c}}},
  {Vanzella}, {Scarlata}, {Boyett}, {Trenti}, \& {Wang}}]{Roy2023}
{Roy}, N., {Henry}, A., {Treu}, T., {et~al.} 2023, \apjl, 952, L14

\bibitem[{{Sanders} {et~al.}(2024){Sanders}, {Shapley}, {Topping}, {Reddy}, \&
  {Brammer}}]{Sanders2023}
{Sanders}, R.~L., {Shapley}, A.~E., {Topping}, M.~W., {Reddy}, N.~A., \&
  {Brammer}, G.~B. 2024, \apj, 962, 24

\bibitem[{{Santini} {et~al.}(2023){Santini}, {Fontana}, {Castellano},
  {Leethochawalit}, {Trenti}, {Treu}, {Belfiori}, {Birrer}, {Bonchi}, {Merlin},
  {Mason}, {Morishita}, {Nonino}, {Paris}, {Polenta}, {Rosati}, {Yang},
  {Boyett}, {Bradac}, {Calabr{\`o}}, {Dressler}, {Glazebrook}, {Marchesini},
  {Mascia}, {Nanayakkara}, {Pentericci}, {Roberts-Borsani}, {Scarlata},
  {Vulcani}, \& {Wang}}]{Santini2023}
{Santini}, P., {Fontana}, A., {Castellano}, M., {et~al.} 2023, \apjl, 942, L27

\bibitem[{{Saxena} {et~al.}(2024){Saxena}, {Bunker}, {Jones}, {Stark},
  {Cameron}, {Witstok}, {Arribas}, {Baker}, {Baum}, {Bhatawdekar}, {Bowler},
  {Boyett}, {Carniani}, {Charlot}, {Chevallard}, {Curti}, {Curtis-Lake},
  {Eisenstein}, {Endsley}, {Hainline}, {Helton}, {Johnson}, {Kumari}, {Looser},
  {Maiolino}, {Rieke}, {Rix}, {Robertson}, {Sandles}, {Simmonds}, {Smit},
  {Tacchella}, {Williams}, {Willmer}, \& {Willott}}]{Saxena2024}
{Saxena}, A., {Bunker}, A.~J., {Jones}, G.~C., {et~al.} 2024, \aap, 684, A84

\bibitem[{{Stiavelli} {et~al.}(2023){Stiavelli}, {Morishita}, {Chiaberge},
  {Grillo}, {Leethochawalit}, {Rosati}, {Schuldt}, {Trenti}, \&
  {Treu}}]{Stiavelli2024}
{Stiavelli}, M., {Morishita}, T., {Chiaberge}, M., {et~al.} 2023, \apjl, 957,
  L18

\bibitem[{{Tang} {et~al.}(2023){Tang}, {Stark}, {Chen}, {Mason}, {Topping},
  {Endsley}, {Senchyna}, {Plat}, {Lu}, {Whitler}, {Robertson}, \&
  {Charlot}}]{Tang2023}
{Tang}, M., {Stark}, D.~P., {Chen}, Z., {et~al.} 2023, \mnras, 526, 1657

\bibitem[{{Treu} {et~al.}(2023){Treu}, {Calabr{\`o}}, {Castellano},
  {Leethochawalit}, {Merlin}, {Fontana}, {Yang}, {Morishita}, {Trenti},
  {Dressler}, {Mason}, {Paris}, {Pentericci}, {Roberts-Borsani}, {Vulcani},
  {Boyett}, {Bradac}, {Glazebrook}, {Jones}, {Marchesini}, {Mascia},
  {Nanayakkara}, {Santini}, {Strait}, {Vanzella}, \& {Wang}}]{Treu2023}
{Treu}, T., {Calabr{\`o}}, A., {Castellano}, M., {et~al.} 2023, \apjl, 942, L28

\bibitem[{{Treu} {et~al.}(2022){Treu}, {Roberts-Borsani}, {Bradac}, {Brammer},
  {Fontana}, {Henry}, {Mason}, {Morishita}, {Pentericci}, {Wang}, {Acebron},
  {Bagley}, {Bergamini}, {Belfiori}, {Bonchi}, {Boyett}, {Boutsia},
  {Calabr{\'o}}, {Caminha}, {Castellano}, {Dressler}, {Glazebrook}, {Grillo},
  {Jacobs}, {Jones}, {Kelly}, {Leethochawalit}, {Malkan}, {Marchesini},
  {Mascia}, {Mercurio}, {Merlin}, {Nanayakkara}, {Nonino}, {Paris},
  {Poggianti}, {Rosati}, {Santini}, {Scarlata}, {Shipley}, {Strait}, {Trenti},
  {Tubthong}, {Vanzella}, {Vulcani}, \& {Yang}}]{Treu2022}
{Treu}, T., {Roberts-Borsani}, G., {Bradac}, M., {et~al.} 2022, \apj, 935, 110

\bibitem[{{Vanzella} {et~al.}(2022){Vanzella}, {Castellano}, {Bergamini},
  {Treu}, {Mercurio}, {Scarlata}, {Rosati}, {Grillo}, {Acebron}, {Caminha},
  {Nonino}, {Nanayakkara}, {Roberts-Borsani}, {Bradac}, {Wang}, {Brammer},
  {Strait}, {Vulcani}, {Me{\v{s}}tri{\'c}}, {Meneghetti}, {Calura}, {Henry},
  {Zanella}, {Trenti}, {Boyett}, {Morishita}, {Calabr{\`o}}, {Glazebrook},
  {Marchesini}, {Birrer}, {Yang}, \& {Jones}}]{Vanzella2022}
{Vanzella}, E., {Castellano}, M., {Bergamini}, P., {et~al.} 2022, \apjl, 940,
  L53

\bibitem[{{Vulcani} {et~al.}(2023){Vulcani}, {Treu}, {Calabr{\`o}}, {Fritz},
  {Poggianti}, {Bergamini}, {Bonchi}, {Boyett}, {Caminha}, {Castellano},
  {Dressler}, {Fontana}, {Glazebrook}, {Grillo}, {Malkan}, {Mascia},
  {Mercurio}, {Merlin}, {Metha}, {Morishita}, {Nanayakkara}, {Paris},
  {Roberts-Borsani}, {Rosati}, {Roy}, {Santini}, {Trenti}, {Vanzella}, \&
  {Wang}}]{Vulcani2023}
{Vulcani}, B., {Treu}, T., {Calabr{\`o}}, A., {et~al.} 2023, \apjl, 948, L15

\bibitem[{{Wang} {et~al.}(2023){Wang}, {Fujimoto}, {Labb{\'e}}, {Furtak},
  {Miller}, {Setton}, {Zitrin}, {Atek}, {Bezanson}, {Brammer}, {Leja}, {Oesch},
  {Price}, {Chemerynska}, {Cutler}, {Dayal}, {van Dokkum}, {Goulding},
  {Greene}, {Fudamoto}, {Khullar}, {Kokorev}, {Marchesini}, {Pan}, {Weaver},
  {Whitaker}, \& {Williams}}]{Wang2023}
{Wang}, B., {Fujimoto}, S., {Labb{\'e}}, I., {et~al.} 2023, \apjl, 957, L34

\bibitem[{{Wang} {et~al.}(2022){Wang}, {Jones}, {Vulcani}, {Treu}, {Morishita},
  {Roberts-Borsani}, {Malkan}, {Henry}, {Brammer}, {Strait}, {Brada{\v{c}}},
  {Boyett}, {Calabr{\`o}}, {Castellano}, {Fontana}, {Glazebrook}, {Kelly},
  {Leethochawalit}, {Marchesini}, {Santini}, {Trenti}, \& {Yang}}]{Wang2022}
{Wang}, X., {Jones}, T., {Vulcani}, B., {et~al.} 2022, \apjl, 938, L16

\bibitem[{{Williams} {et~al.}(2023){Williams}, {Kelly}, {Chen}, {Brammer},
  {Zitrin}, {Treu}, {Scarlata}, {Koekemoer}, {Oguri}, {Lin}, {Diego}, {Nonino},
  {Hjorth}, {Langeroodi}, {Broadhurst}, {Rogers}, {Perez-Fournon}, {Foley},
  {Jha}, {Filippenko}, {Strolger}, {Pierel}, {Poidevin}, \&
  {Yang}}]{Williams2023}
{Williams}, H., {Kelly}, P.~L., {Chen}, W., {et~al.} 2023, Science, 380, 416

\bibitem[{{Witstok} {et~al.}(2024){Witstok}, {Smit}, {Saxena}, {Jones},
  {Helton}, {Sun}, {Maiolino}, {Kumari}, {Stark}, {Bunker}, {Arribas}, {Baker},
  {Bhatawdekar}, {Boyett}, {Cameron}, {Carniani}, {Charlot}, {Chevallard},
  {Curti}, {Curtis-Lake}, {Eisenstein}, {Endsley}, {Hainline}, {Ji}, {Johnson},
  {Looser}, {Nelson}, {Perna}, {Rix}, {Robertson}, {Sandles}, {Scholtz},
  {Simmonds}, {Tacchella}, {{\"U}bler}, {Williams}, {Willmer}, \&
  {Willott}}]{Witstok2024}
{Witstok}, J., {Smit}, R., {Saxena}, A., {et~al.} 2024, \aap, 682, A40

\bibitem[{{Yang} {et~al.}(2022){Yang}, {Morishita}, {Leethochawalit},
  {Castellano}, {Calabr{\`o}}, {Treu}, {Bonchi}, {Fontana}, {Mason}, {Merlin},
  {Paris}, {Trenti}, {Roberts-Borsani}, {Bradac}, {Vanzella}, {Vulcani},
  {Marchesini}, {Ding}, {Nanayakkara}, {Birrer}, {Glazebrook}, {Jones},
  {Boyett}, {Santini}, {Strait}, \& {Wang}}]{Yang2022}
{Yang}, L., {Morishita}, T., {Leethochawalit}, N., {et~al.} 2022, \apjl, 938,
  L17

\bibitem[{{Zavala} {et~al.}(2024){Zavala}, {Castellano}, {Akins}, {Bakx},
  {Burgarella}, {Casey}, {Ch{\'a}vez Ortiz}, {Dickinson}, {Finkelstein},
  {Mitsuhashi}, {Nakajima}, {P{\'e}rez-Gonz{\'a}lez}, {Arrabal Haro}, {Buat},
  {Backhaus}, {Calabr{\`o}}, {Cleri}, {Fern{\'a}ndez-Arenas}, {Fontana},
  {Franco}, {Giavalisco}, {Grogin}, {Hathi}, {Hirschmann}, {Ikeda}, {Jung},
  {Kartaltepe}, {Koekemoer}, {Larson}, {McKinney}, {Papovich}, {Saito},
  {Santini}, {Terlevich}, {Terlevich}, {Treu}, \& {Yung}}]{Zavala2024}
{Zavala}, J.~A., {Castellano}, M., {Akins}, H.~B., {et~al.} 2024, arXiv
  e-prints, arXiv:2403.10491

\end{thebibliography}
\bibliographystyle{aa}

\onecolumn
\newpage 

\begin{multicols}{2}

\section*{Appendix A: Spectroscopic catalog}
This appendix includes the list of the detected emission lines (Table A.\ref{tab:spectral_lines}).

\end{multicols}

\begin{longtable}{llc}
\caption*{Table A.1: Observed spectral emission lines for the GLASS-JWST and the JWST-DD-2756 spectra. Doublets are marked by (1), (2).} \label{tab:spectral_lines}\\
\hline\hline
Ion & Notation & Rest-frame $\lambda$ (\AA)\\
\hline
\endfirsthead
\caption*{Table A.1: continued.}\\
\hline\hline
Ion & Notation & Rest-frame $\lambda$ (\AA)\\
\hline
\endhead
\hline
\endfoot
\text{Ly}$\alpha$ & H1[1216] & 1215.670\\
\text{C IV} & C4[1548] & 1548.187\\ 
\text{He II} & He2[1640] & 1640.391\\ 
\text{O III]} & O3[1666] & 1666.150	\\
\text{C III]} & C3[1909] & 1908.734\\
\text{Mg II} & Mg2[2803] & 2803.531\\
\text{[Ne V]} & Ne5[3426] &3425.881\\
\text{H16} & H1[3704] & 3704.913\\
\text{[O II] (1)} &O2[3727]& 3727.100\\
\text{[O II] (2)} & O2[3729] & 3729.860\\
\text{H12} & H1[3750] & 3751.224\\
\text{H11} & H1[3771] & 3771.708\\
\text{H10} & H1[3798] & 3798.982\\
\text{H9} & H1[3835] & 3836.479\\
\text{[Ne III]} & Ne3[3869] & 3869.860\\
\text{H8} & H1[3889] & 3890.158\\ 
\text{H}$\epsilon$ & H1[3970] & 3971.202\\ 
\text{He I} & He1[4026]& 4027.335\\
\text{[S II]}& S2[4069] & 4069.749\\
\text{H}$\delta$ & H1[4102] & 4102.899\\ 
\text{H}$\gamma$ & H1[4340] & 4341.691\\ 
\text{[O III]} & O3[4363] & 4364.436\\
\text{He I} & He1[4471] & 4472.740\\
\text{He II} & He2[4685]& 4686.879\\
\text{[Fe III]} & Fe3[4658] & 4659.400\\
\text{[Ar IV] (1)} & Ar4[4711] & 4712.580\\
\text{[Ar IV] (2)} & Ar4[4740] & 4741.450\\
\text{H}$\beta$ & H1[4861] & 4862.691\\ 
\text{He I} & He1[4922] & 4923.305\\
\text{[O III] (1)} & O3[4959] &4960.295\\ 
\text{[O III] (2)} & O3[5007] & 5008.240\\ 
\text{[N II]} & N2[5755] & 5755.000 \\
\text{He I} & He1[5876] & 5877.243\\
\text{[O I]} & O1[6300] & 6302.046\\
\text{[S III]} & S3[6312] & 6313.810\\
\text{H$\alpha$} & H1[6563] & 6564.632\\
\text{He I} & He1[6678] & 6679.996\\
\text{[S II] (1)} & S2[6716] & 6718.295\\
\text{[S II] (2)} & S2[6731] & 6732.674\\
\text{He I} & He1[7065] & 7067.163\\
\text{[Ar III] (1)} & Ar3[7136] & 7137.760\\ 
\text{[O II] (1)} & O2[7319] & 7320.940 \\
\text{[O II] (2)}& O2[7330] & 7331.680 \\
\text{[Ar III] (2)} & ]r3[7751] & 7753.240\\ 
\text{Pa20} & H1[8392] & 8394.703 \\
\text{Pa19} & H1[8413] & 8415.630\\
\text{Pa18} & H1[8438] & 8440.274 \\
\text{Pa17} & H1[8467]& 8469.581	\\
\text{Pa16} & H1[8502] & 8504.819 \\
\text{Pa15} & H1[8545] & 8547.731\\
\text{Pa14} & H1[8598] & 8600.754	\\
\text{Pa13} & H1[8665] & 8667.398 \\
\text{Pa12} & H1[8750] & 8752.876 \\
\text{Pa11} & H1[8863] & 8865.216 \\ 
\text{Pa10} & H1[9015] & 9017.384 \\
\text{[S III] (1)} & S3[9068] & 9071.100\\ 
\text{Pa9} & H1[9229] & 9231.546 \\ 
\text{[S III] (2)} & S3[9530] &  9533.200\\
\text{Pa}$\epsilon$ & H1[9546] & 9548.588 \\ 
\text{Pa}$\delta$ & H1[10049] & 10052.123 \\ 
\text{He I} & He1[10830] & 10833.306 \\
\text{Pa}$\gamma$ & H1[10938] & 10941.082\\ 
\text{[O I]} & O1[11287] & 11287.000\\
\text{Pa}$\beta$ & H1[12818] & 12821.576\\ 
\text{Fe II} & Fe2[16443] & 16447.955\\ 
\text{Pa}$\alpha$ & H1[18751] & 18756.096\\
\end{longtable}

\end{document}